\documentclass[a4paper,fleqn]{cas-dc}

\usepackage[authoryear]{natbib}

\usepackage{graphicx}
\usepackage{tikz} 

\usepackage{subfigure}
\usepackage{epstopdf}
\usepackage{color}
\usepackage[noend]{algpseudocode}
\usepackage{algorithmicx,algorithm}

\usepackage{url}            % simple URL typesetting
\usepackage{booktabs}       % professional-quality tables
\usepackage{amsfonts}       % blackboard math symbols
\usepackage{nicefrac}       % compact symbols for 1/2, etc.
\usepackage{microtype}      % microtypography
\usepackage{enumitem}
\usepackage{bm}
\usepackage{nccmath}
\usepackage{boldline}
\usepackage{booktabs} % To thicken table lines

\usepackage{algorithm}
\usepackage[noend]{algcompatible}

\usepackage{capt-of}
\usepackage{color,soul}

\usepackage{epsfig}
\usepackage{multirow}
\usepackage{xcolor}
\usepackage{amsmath}
\usepackage{amssymb}
\usepackage{wrapfig}
\usepackage{comment}

\usepackage{textcomp}
\usepackage{soul}

\usepackage{caption}
\usepackage{times}  % 或使用 \usepackage{newtxtext}
\usepackage{makecell}
\usepackage{subfigure}
\usepackage{arydshln}
\usepackage{ulem}
\usepackage{marginnote} % <-------------
\usepackage{url}
\usepackage{bbm}

\usepackage[misc]{ifsym} 
\usepackage{mathtools}

\usepackage{verbatim}
\usepackage{float}

\usepackage{bbding}

\useunder{\underline}{\ul}{}
\usepackage{xcolor}
\usepackage{tikz}

\def\tsc#1{\csdef{#1}{\textsc{\lowercase{#1}}\xspace}}
\tsc{WGM}
\tsc{QE}
\tsc{EP}
\tsc{PMS}
\tsc{BEC}
\tsc{DE}
\usepackage{hyperref}
\begin{document}
\begin{sloppypar}
	\let\WriteBookmarks\relax
	\def\floatpagepagefraction{1}
	\def\textpagefraction{.001}
	\let\printorcid\relax
	\shorttitle{}
	\shortauthors{Yang et~al.} %% 缩略作者 自己名字， 比如： 张三 = S. Zhang

	%% 标题
	\title [mode = title]{\textsc{MM-InstructEval}: Zero-Shot Evaluation of (Multimodal) Large Language Models on Multimodal Reasoning Tasks}
        
	%%\tnotemark[1,2]

	%%\tnotetext[1]{This document is the results of the research project funded by the National Science Foundation.}

	%%\tnotetext[2]{The second title footnote which is a longer text matter to fill through the whole text width and overflow into another line in the footnotes area of the first page.}

	%% 作者顺序
	%% 1
      
	\author[1,2]{\textcolor[RGB]{0,0,1}{Xiaocui Yang}}
	\ead{yangxiaocui@cse.neu.edu.cn}
    
	%% 2
	\author[1,3]{\textcolor[RGB]{0,0,1}{Wenfang Wu}}
        \ead{wenfang@stumail.neu.edu.cn}
	% \fnmark[1] %%第几作者
    
        \author[1]{\textcolor[RGB]{0,0,1}{Shi Feng}}
        \ead{fengshi@cse.neu.edu.cn}
        
        \author[1]{\textcolor[RGB]{0,0,1}{Ming Wang}}
        \ead{sci.m.wang@gmail.com}
        
        \author[1]{\textcolor[RGB]{0,0,1}{Daling Wang}}
        \ead{wangdaling@cse.neu.edu.cn}
        
        \author[1]{\textcolor[RGB]{0,0,1}{Yang Li}}
        \ead{liyang@stumail.neu.edu.cn}
        
        \author[2]{\textcolor[RGB]{0,0,1}{Qi Sun}}
        \ead{319106003718@njust.edu.cn}
        
        \author[1]{\textcolor[RGB]{0,0,1}{Yifei Zhang}}
        \ead{zhangyifei@cse.neu.edu.cn}
        
        \author[3]{\textcolor[RGB]{0,0,1}{Xiaoming Fu}}
        \ead{fu@cs.uni-goettingen.de}
        
        \author[2]{\textcolor[RGB]{0,0,1}{Soujanya Poria}}
        \ead{sporia@sutd.edu.sg}
        \cormark[1]%%通讯作者星标
    
	%\credit{}%%本文的贡献
	\address[1]{School of Computer Science and Engineering, Northeastern University, Shenyang, China}
	\address[2]{Singapore University of Technology and Design, Singapore}
        \address[3]{University of Göttingen, Germany}

	\cortext[cor1]{Corresponding author.} %% 首页左下角通讯作者
	%%\cortext[cor2]{Principal corresponding author} 

	% \fntext[fn1]{Equal Contribution.}
	%%\fntext[fn2]{Another author footnote, this is a very long footnote and it should be a really long footnote. But this footnote is not yet sufficiently long enough to make two lines of footnote text.}

	%%\nonumnote{This note has no numbers. In this work we demonstrate $a_b$ the formation Y\_1 of a new type of polariton on the interface between a cuprous oxide slab and a polystyrene micro-sphere placed on the slab.}
       
	%%摘要
	\begin{abstract}
       The emergence of multimodal large language models (MLLMs) has triggered extensive research in model evaluation. While existing evaluation studies primarily focus on unimodal (vision-only) comprehension and reasoning capabilities, they overlook critical assessments of complex multimodal reasoning tasks that require integrated understanding of both visual and textual contexts. Such multimodal tasks present unique challenges, demanding sophisticated reasoning across multiple modalities and deep comprehension of multimodal contexts.
        In this paper, we present \textbf{\textsc{MM-InstructEval}}, a comprehensive evaluation framework that incorporates diverse metrics to assess model performance across various multimodal reasoning tasks with vision-text contexts. We conduct extensive zero-shot evaluations on 45 models (including 36 MLLMs) across 16 multimodal datasets, encompassing 6 distinct tasks using 10 different instructions. Our framework introduces multiple innovative metrics, including the ``Best Performance'' metric to benchmark peak model capabilities, the ``Mean Relative Gain'' metric to assess overall efficacy across models and instructions, the ``Stability'' metric to measure robustness, and the ``Adaptability'' metric to quantify the compatibility between models and instructions.
        Through comprehensive evaluation and analysis, we uncover several significant insights about model architectures, instruction formats, and their interactions in multimodal reasoning tasks. Our findings establish new benchmarks for assessing the reasoning capabilities of MLLMs and provide strategic guidance for future developments. To facilitate continued research and evaluation in this field, we release our framework and resources at \url{https://github.com/declare-lab/MM-InstructEval}, with an interactive leaderboard available at \href{https://declare-lab.github.io/MM-InstructEval/}{\textsc{MM-InstructEval} Leaderboard}.
        %增加了10个模型，8个开源的 MLLM， 2个闭源的

	\end{abstract}

	% 	%%\includegraphics{./grabs.pdf} %%图片摘要地址路径

	%%高亮
	% \begin{highlights}
	% 	\item End-to-end community detection method based on graph convolution network.
	% 	\item A new community perspective similarity is proposed.
	% 	\item Modify the convolution layer for large networks.
	% 	\item The loss function based on modularity and Bernoulli Poisson model is introduced.
	% 	\item Evaluate performance using real-world networks.
	% \end{highlights}
		
	%% 关键词

	% 此指令为生成标题格式，不可删除
	\maketitle
        \begin{tikzpicture}[remember picture,overlay,shift={(current page.north west)}]
        \node[anchor=north west,xshift=0.75cm,yshift=-1.75cm]{\scalebox{-1}[1]{\includegraphics[width=0.8cm]{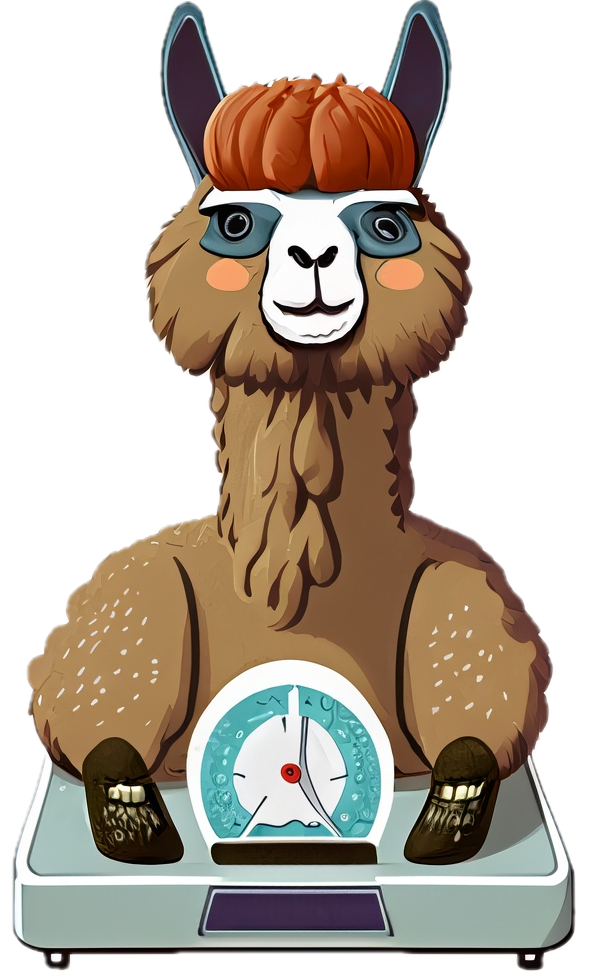}}};
        \end{tikzpicture}

	%% 1.引言
	\section{Introduction}
	%%\par{文本内容}换行并缩进
     %    \par{
    	% 	Many approaches have been proposed.
    	% }
        \begin{figure*}[t] %%
          \centering %?????????
          \includegraphics[width = 1\textwidth]{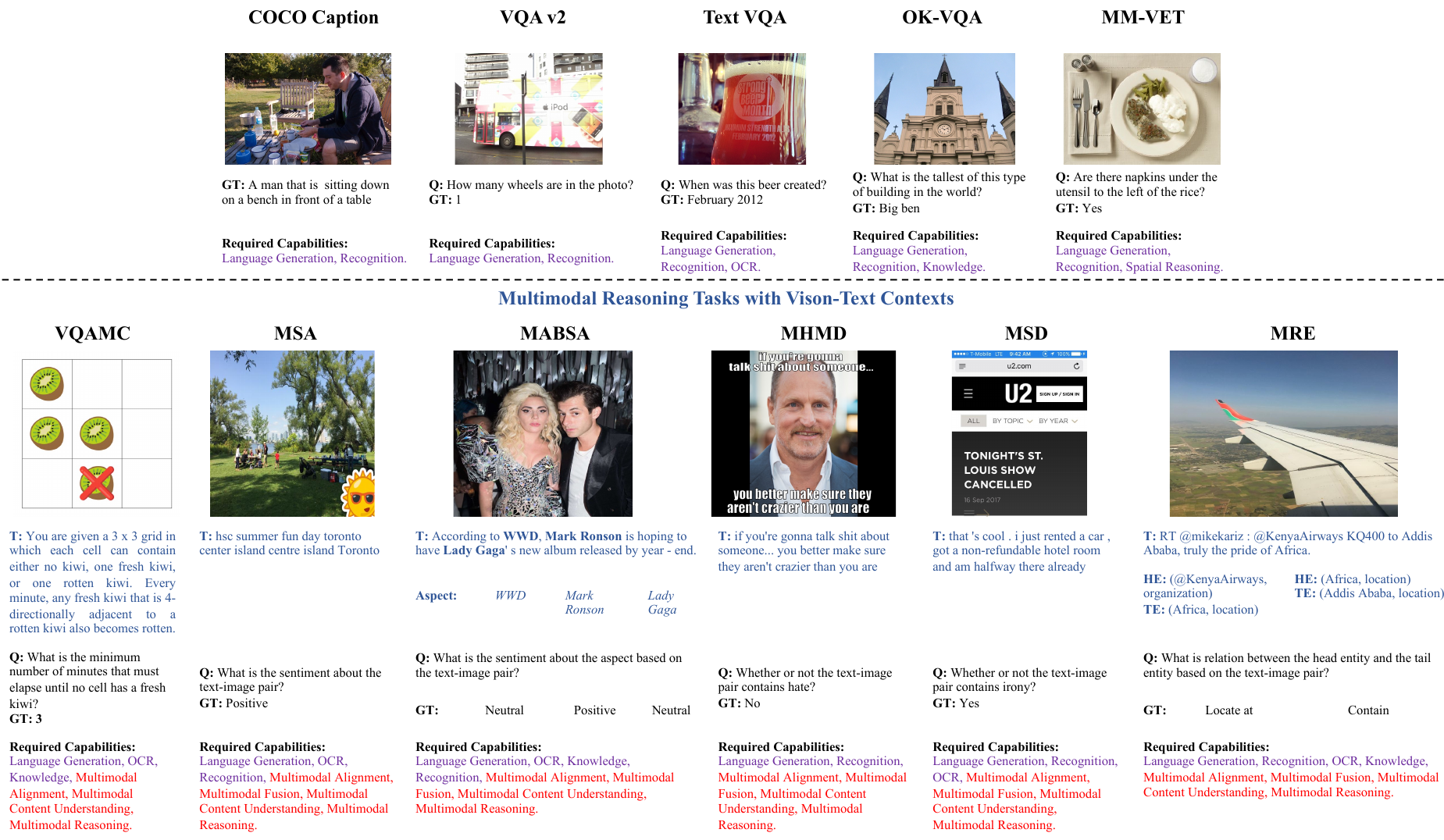}
          %\vpsace{-1em}       
          \caption{Required capabilities for diverse datasets.
          Different from the \textcolor[RGB]{153,50,204}{traditional capabilities} indicated above the dotted line for multimodal reasoning tasks with vision-only contexts—including COCO \citep{DBLP:conf/eccv/LinMBHPRDZ14}, VQA v2 \citep{DBLP:conf/eccv/LinMBHPRDZ14}, Text VQA \citep{DBLP:conf/cvpr/SinghNSJCBPR19}, OK-VQA \citep{DBLP:conf/cvpr/MarinoRFM19}, and MM-VET \citep{DBLP:journals/corr/abs-2308-02490}—our own multimodal reasoning tasks that incorporate vision-text contexts, particularly those below the dotted line, require not only capabilities relevant to traditional tasks but also a profound  \textcolor{red}{interaction and understanding} of complex vision-text contexts. In these tasks, `T' represents the text context, `Q' denotes the question prompting the models for answers, and `GT' stands for the ground truth label. `HE' and `TE' correspond to the head entity and tail entity, respectively.
           }
          \label{Figure_multimodal reasoning Tasks} 
          %\vpsace{-1.5em}
        \end{figure*}

        Multimodal Large Language Models (MLLMs), harnessing the formidable capabilities of Large Language Models (LLMs), demonstrate outstanding performance across a spectrum of multimodal tasks \citep{alayrac2022flamingo, DBLP:journals/corr/abs-2304-15010, DBLP:journals/corr/abs-2301-12597, dai2023instructblip}. 
        Recent research developments, as summarized in Table \ref{tab:other_benchmarks}, include but are not limited to MME \citep{DBLP:journals/corr/abs-2306-13394}, MMBench \citep{liu2023mmbench}, SEED-Bench \citep{DBLP:journals/corr/abs-2307-16125}, LVLM-eHub \citep{DBLP:journals/corr/abs-2306-09265}, MM-Vet \citep{DBLP:journals/corr/abs-2308-02490}, MMMU \citep{yue2023mmmu}, and MLLM-Bench \citep{ge2023mllm}. These studies primarily focus on evaluating the traditional vision-language multimodal capabilities of MLLMs in tasks predominantly driven by visual context (vision context + text question), such as Visual Question Answering (VQA) and Video Question Answering (VideoQA). In these tasks, the answers to the questions are often derived solely based on the image context,  as illustrated in the section above the dotted line in Figure \ref{Figure_multimodal reasoning Tasks}.
        These capabilities include Recognition, Optical Character Recognition (OCR), Spatial Reasoning, Knowledge Integration, and Code Reasoning \citep{DBLP:journals/corr/abs-2308-02490}. However, the performance of Multimodal Large Language Models (MLLMs) in multimodal reasoning tasks that combine vision-text contexts (vision-\textbf{\textcolor[RGB]{70,130,180}{text contexts}} + text question) remains less explored. Such tasks include Visual Question Answering with Multimodal Contexts (VQAMC) \citep{DBLP:conf/nips/LuMX0CZTCK22, yue2023mmmu, chia-etal-2024-puzzlevqa}, Multimodal Sentiment Analysis (MSA) \citep{DBLP:conf/mmm/NiuZPE16, DBLP:journals/tmm/YangFW021, DBLP:journals/corr/ZadehZPM16, DBLP:conf/acl/MorencyCPLZ18}, Multimodal Aspect-Based Sentiment Analysis (MABSA) \citep{DBLP:conf/aaai/0001FLH18, DBLP:conf/acl/JiZCLN18, DBLP:journals/ijon/ZhouZHHH21}, Multimodal Hateful Memes Detection (MHMD) \citep{mathias2021findings}, Multimodal Sarcasm Detection (MSD) \citep{DBLP:conf/acl/CaiCW19}, and Multimodal Relation Extraction (MRE) \citep{mathias2021findings}.
        In the field of Natural Language Processing (NLP), most research \citep{DBLP:journals/corr/abs-2305-15005, DBLP:journals/corr/abs-2304-04339, DBLP:journals/corr/abs-2308-04945} primarily evaluates pure Large Language Models (LLMs) like ChatGPT \citep{chatgpt}, Flan-T5 \citep{DBLP:journals/corr/abs-2210-11416}, focusing specifically on text classification tasks such as text sentiment analysis and relation classification. It leaves a gap in understanding the performance of various MLLMs in multimodal reasoning tasks that rely on both text and image contexts.
        For example, in the Visual Question Answering with Multimodal Contexts (VQAMC) task depicted in Figure \ref{Figure_multimodal reasoning Tasks}, it is clear that the answer to the question depends on integrating the image-text contexts and multi-step reasoning. Simply using the image and the question without incorporating the textual context fails to produce the correct answer.
         
        To bridge this research gap, we present a comprehensive evaluation of 45 publicly available models, including 36 MLLMs, across 16 datasets encompassing 6 distinct tasks. These tasks demand not only traditional vision-language capabilities but also sophisticated reasoning abilities to integrate and understand multimodal contexts effectively. The complexity of such tasks is exemplified in VQAMC, where correct responses require the integration of both visual and textual information, as simple image-question pairs without textual context prove insufficient. Our evaluation framework addresses the multifaceted nature of multimodal reasoning with vision-text contexts, which encompasses various cognitive processes, including multimodal alignment, interaction, and fusion \citep{DBLP:conf/emnlp/HanCP21, DBLP:conf/mm/HazarikaZP20, wu2023sentimental}, alongside content understanding \citep{wu2024transferring}. While defining the boundaries of such reasoning can be challenging due to the diverse elements involved (e.g., visual-language alignment, OCR, spatial reasoning), datasets like AlgoPuzzleVQA \citep{chia-etal-2024-puzzlevqa} demonstrate how these reasoning capabilities are activated through carefully designed prompts and instructions. 

        \begin{table*}[t]
        \centering
        \small
        \setlength{\tabcolsep}{2pt}
        \renewcommand{\arraystretch}{1.1} 
        \caption{Comparison of our proposed \textbf{\textsc{MM-InstructEval}} with recent vision-language benchmarks.
        Traditional capabilities, as mentioned in previous studies \citep{DBLP:journals/corr/abs-2308-02490, DBLP:journals/corr/abs-2307-16125}, include Language Generation, Recognition, Knowledge Reasoning, Spatial Reasoning, OCR, and more. 
        Multimodal reasoning tasks with multimodal contexts more concentrate on Multimodal Alignment, Multimodal Fusion, Multimodal Content Understanding, and other abilities. 
        \textsc{MM-InstructEval} also conducted a study about the adaptability between diverse models and instructions.
        }
        \resizebox{1\textwidth}{!}{
        \begin{tabular}{lccccccc}
        \toprule
        \textbf{Benchmark} & \textbf{Purpose} &  \textbf{Text Context} & \textbf{Image} & \textbf{Question} & \textbf{Answer Type} & \textbf{Evaluator} \\
        \midrule
        MME~\citep{DBLP:journals/corr/abs-2306-13394} & Traditional Capabilities & \color{purple}{\XSolidBrush} & \color{teal}{\Checkmark} & \color{teal}{\Checkmark}  & Yes/No & Metrics \\
        MMBench~\citep{liu2023mmbench} & Traditional Capabilities & \color{purple}{\XSolidBrush} & \color{teal}{\Checkmark} & \color{teal}{\Checkmark}  & Multi-choice & GPT \\
        SEED-Bench~\citep{DBLP:journals/corr/abs-2307-16125}  & Traditional Capabilities & \color{purple}{\XSolidBrush} & \color{teal}{\Checkmark} & \color{teal}{\Checkmark}  & Multi-choice & Metrics \\
        LVLM-eHub~\citep{DBLP:journals/corr/abs-2306-09265}  & Traditional Capabilities & \color{purple}{\XSolidBrush} & \color{teal}{\Checkmark} & \color{teal}{\Checkmark}  &  Multi-choice/Open-ended & Metrics \\
        MM-Vet~\citep{DBLP:journals/corr/abs-2308-02490} & Traditional Capabilities & \color{purple}{\XSolidBrush} & \color{teal}{\Checkmark} & \color{teal}{\Checkmark}  & Multi-choice/Open-ended & Metrics/GPT-4 \\
        MLLM-Bench~\citep{ge2023mllm} & Traditional Capabilities & \color{purple}{\XSolidBrush} & \color{teal}{\Checkmark} & \color{teal}{\Checkmark}  & Open-ended & GPT-4V/LLaVA \\
        MMMU~\citep{yue2023mmmu}  & Multimodal Reasoning & \color{teal}{\Checkmark} & \color{teal}{\Checkmark} & \color{teal}{\Checkmark}  &  Multi-choice/Open-ended & Metrics \\ 
        \midrule
        HallusionBench~\citep{guan2023hallusionbench} & Visual Hallucination  & \color{purple}{\XSolidBrush} & \color{teal}{\Checkmark} & \color{teal}{\Checkmark} & Yes/No & Metrics \\
        Bingo~\citep{cui2023holistic} & Visual Hallucination &  \color{purple}{\XSolidBrush} & \color{teal}{\Checkmark} & \color{teal}{\Checkmark} & Open-ended & Human \\
        \midrule
        \textbf{\textsc{MM-InstructEval}} & Multimodal Reasoning/Instructions & \color{teal}{\Checkmark} & \color{teal}{\Checkmark} & \color{teal}{\Checkmark} &  Multi-choice & Metrics \\
        \bottomrule[1pt]
        \end{tabular}}
        
        \label{tab:other_benchmarks}
        \end{table*}

\begin{figure*}[t] %%
          \centering %?????????
          \includegraphics[width = 1\textwidth]{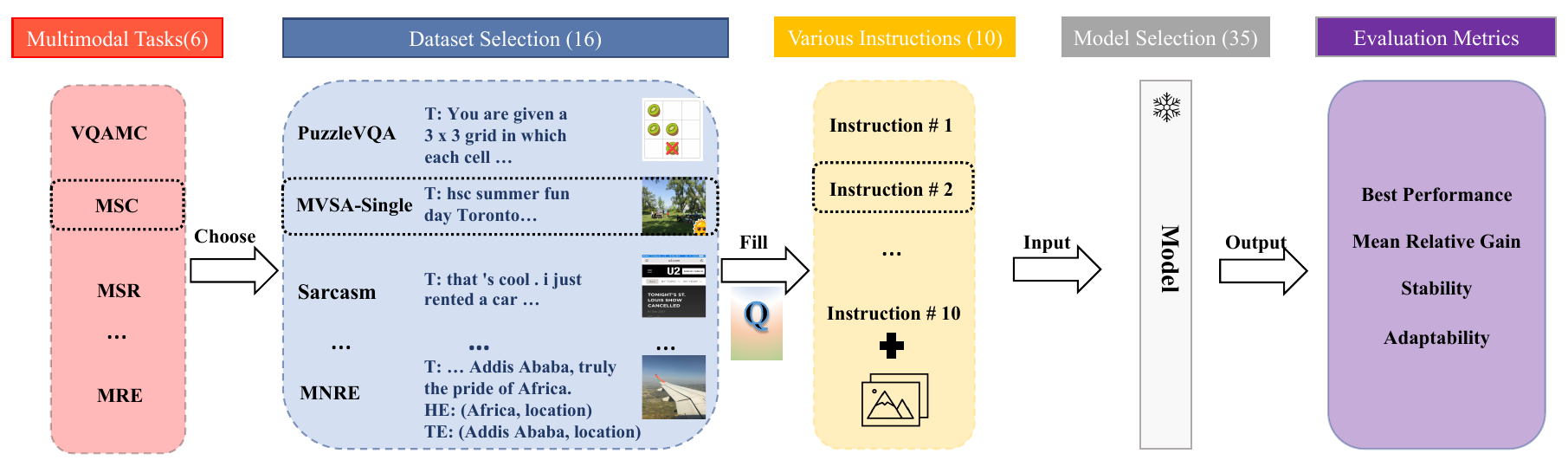}
          %\vpsace{-1em}
          \caption{Overview of our \textbf{\textsc{MM-InstructEval}} framework, which conducts evaluations of popular MLLMs across various multimodal reasoning tasks with multimodal contexts utilizing comprehensive metrics. As illustrated in the dashed box, we select the `MVSA-Single' dataset from the `MSC' task to utilize `Instruction \# 2' for evaluating a specific `MLLM'. We then aggregate results to thoroughly assess the performance of models and instructions using a variety of metrics.
          A colorful `Q' symbolizes `Question, ' and its design varies according to the specific task. For more detailed visual representations and explanations, please refer to Figure \ref{Figure_2_text_instruction_for_different_tasks}.
           }
          \label{Figure_MM-InstructEval-Framework} 
          %\vpsace{-1.5em}
        \end{figure*}

        We present \textbf{\textsc{MM-InstructEval}}, a comprehensive assessment framework that integrates diverse metrics to evaluate models and instructions in multimodal reasoning tasks with vision-text contexts (Figure \ref{Figure_MM-InstructEval-Framework}). This framework complements existing zero-shot evaluations of MLLMs, offering a more holistic assessment approach through several innovative metrics: The ``Best Performance'' metric establishes the peak capabilities of each model under optimal instructions for specific datasets. To capture overall effectiveness, we introduce the ``Model Mean Relative Gain'' metric for assessing model performance across instructions, and the ``Instruction Mean Relative Gain'' metric for evaluating instruction effectiveness across models. The ``Model Stability'' and ``Instruction Stability'' metrics measure robustness by evaluating consistency of performance across different scenarios. While previous studies have focused separately on either model evaluation \citep{DBLP:journals/corr/abs-2306-13394, DBLP:journals/corr/abs-2307-16125, DBLP:journals/corr/abs-2306-09265, DBLP:journals/corr/abs-2308-02490, yue2023mmmu, ge2023mllm} or instruction assessment \citep{DBLP:journals/corr/abs-2306-04757, DBLP:journals/corr/abs-2307-00259}, we propose the ``Adaptability'' metric to quantify how well different models adapt to various instructions. This metric measures how frequently instructions achieve Top-K performance across datasets for specific models, providing crucial insights into optimal instruction design for different models. 
        
        Our extensive experiments reveal significant insights into the performance of large (multimodal) language models in multimodal reasoning tasks with vision-text contexts: First, advanced closed-source models (GPT-4V, GLM-4V-plus) consistently outperform open-source alternatives across most tasks, with particularly notable performance gaps emerging in challenging scenarios like multimodal relation extraction and the VQAMA task requiring multi-step reasoning. Within open-source models, Encoder-Decoder architectures (using Flan-T5 backbone, e.g., BLIP-2, InstructBLIP) demonstrate superior performance compared to the LLaMA series. Second, the latest models, such as Qwen2.5-VL, MiciCPM-V2.6, and others, showcase exceptional performance through architectural innovations and improved training strategies. Notably, newer models achieve comparable results with significantly fewer parameters (e.g., Qwen2.5-VL-7B-Instruct vs. Qwen2-VL-72B-Instruct), indicating higher information density and more effective parameter utilization in latest models. Third, regarding instruction format impact, ``Question-Answer'' format consistently yields superior results, and including structured options within instructions further enhances performance on specific tasks, providing practical guidance for optimizing model interactions in real-world applications. Fourth, different models show varying preferences for instruction formats, reflecting their diverse training paradigms, which emphasizes the importance of flexible instruction strategies for optimal performance and is crucial for maximizing model effectiveness across different tasks. Fifth, while current open-source MLLMs excel in basic vision-text reasoning, they struggle with more complex scenarios, showing significant performance gaps in multi-step reasoning tasks (e.g., PuzzleVQA), fine-grained information extraction, and long-tail phenomena in multimodal relation extraction. These findings provide valuable insights for both theoretical understanding and practical applications of MLLMs, while also identifying key areas for future development in complex reasoning capabilities and model architecture design.

        Our main contributions are summarized as follows:
        %\vpsace{-2em}
        \begin{itemize}
        \item We introduce the \textsc{MM-InstructEval} framework, which incorporates multiple metrics for detailed model and instruction evaluation, including the Best Performance metric for benchmarking peak capabilities, the Mean Relative Gain metric for assessing overall efficacy, the Stability metric for measuring robustness, and the Adaptability metric for quantifying model-instruction compatibility, offering valuable insights for instruction design to maximize current and future model performance.
        
        \item We conduct comprehensive zero-shot evaluations of 45 models across 16 datasets, employing 10 distinct instructions to assess 6 diverse multimodal reasoning tasks that require integrated vision-text understanding. This represents the first extensive assessment of MLLMs across various instructions, with a particular focus on complex multimodal reasoning scenarios.
        
        \item 
        Through extensive experimentation, we establish comprehensive benchmarks for both LLMs and MLLMs across various multimodal reasoning tasks with vision-text contexts.
        We significantly expand upon previous MLLM evaluations by introducing systematic comparisons across models, datasets, and instructions. This extensive evaluation uncovers critical insights into model behaviors, limitations, and performance variability, providing practitioners with practical guidance for model selection and instruction design to optimize performance in multimodal tasks.
        
        \end{itemize}

	%% 2.第二章
	\section{Related work}
	\label{Related_work}
	\subsection{LLMs-Based Evaluation}
        As Large Language Models (LLMs) \citep{DBLP:journals/corr/abs-2306-13549} gain prominence, numerous evaluations of these models have surfaced. For example,  \citeauthor{DBLP:journals/corr/abs-2305-15005} conduct comprehensive evaluations of LLMs, including ChatGPT and Flan-T5, across a variety of sentiment analysis tasks, ranging from text sentiment analysis to aspect-based sentiment analysis and multifaceted analysis of subjective texts. Additionally, \citeauthor{DBLP:journals/corr/abs-2304-04339} assess the performance of ChatGPT across five representative sentiment analysis tasks.
        While most of the aforementioned studies primarily focus on text sentiment analysis and evaluate only a limited number of LLMs, there is a growing need for more comprehensive evaluation frameworks. To address this gap, LLMeBench introduces an open-source, user-friendly, and adaptable benchmarking framework designed specifically for LLMs. It features four key components: the Dataset module, Asset module, Model module, and Evaluation module.
        INSTRUCTEVAL \citep{DBLP:journals/corr/abs-2306-04757} introduces a comprehensive evaluation suite specifically designed for 11 instruction-tuned large language models, including models such as Flan-T5, Vicuna, Alpaca, and others. In a related vein, InstructEval \citep{DBLP:journals/corr/abs-2307-00259} methodically explores the generalizability of popular instruction selection and induction methods for in-context learning (ICL) in LLMs.
        
        While these studies are pivotal in evaluating LLMs, they predominantly focus on text-based tasks. Additionally, the aforementioned research tends to concentrate either solely on assessing different models or exclusively on evaluating the effectiveness of instructions, thereby overlooking the critical aspect of adaptability between models and instructions.
        Our primary focus is to assess the performance of various MLLMs across diverse multimodal reasoning tasks with vision-text contexts. To address the noted research gap, we propose the Adaptability metric to quantify the degree of adaptability between different models and various instructions, aiming to foster a deeper understanding of how MLLMs can be optimally utilized in complex, real-world scenarios.
        
        % \vspace{-1em}
        \subsection{MLLMs-Based Evaluation}
        Multimodal Large Language Models (MLLMs) \citep{DBLP:journals/corr/abs-2306-13549}, building upon the capabilities of LLMs, excel in a variety of multimodal tasks such as Caption Generation and Visual Question Answering. The proficiency spurs a surge in research of evaluating these models. We summarize the latest benchmarks for MLLMs in Table \ref{tab:other_benchmarks}. Specifically,
        MME \citep{DBLP:journals/corr/abs-2306-13394} introduces the first MLLM evaluation benchmark, categorizing tasks into Perception (Recognition, OCR, and others) and Cognition (Commonsense Reasoning, Numerical Calculation, Text Translation, and so on).
        MMBench \citep{liu2023mmbench} provides a systematic framework to robustly evaluate the diverse capabilities of large vision-language models.
        SEED-Bench \citep{DBLP:journals/corr/abs-2307-16125} evaluates comprehension across image and video modalities through 19K multiple-choice questions covering 12 dimensions.
        LVLM-eHub \citep{DBLP:journals/corr/abs-2306-09265} assesses 8 MLLMs, including InstructBLIP and MiniGPT, on 47 visual text-related benchmarks, featuring quantitative evaluations and an online arena platform.
        MM-Vet \citep{DBLP:journals/corr/abs-2308-02490} benchmarks MLLMs across 16 tasks, identifying six core Visual Language (VL) capabilities essential for complex multimodal tasks.
        MMMU benchmark \citep{yue2023mmmu} aims to evaluate expert-level multimodal understanding capabilities of foundation models across a broad range of tasks.
        MLLM-Bench \citep{ge2023mllm} utilizes small datasets and the capabilities of GPT-4V \citep{GPT4V} to assess the performance of various open-source MLLMs.
        Additional evaluations like HallusionBench \citep{guan2023hallusionbench} and Bingo \citep{cui2023holistic} focus on the hallucination phenomenon in MLLMs.
        
        Existing benchmarks primarily focus on vision-driven tasks where questions can typically be answered based solely on visual context. However, these benchmarks overlook multimodal reasoning tasks that require the integration of both visual and textual contexts to effectively answer questions. In such tasks, relying solely on visual content is insufficient for deriving correct answers, necessitating a deeper understanding and reasoning across modalities. We expand upon existing research by conducting a more extensive and comprehensive evaluation of MLLMs in diverse reasoning tasks involving complex multimodal contexts. Our evaluation not only enhances the understanding about capabilities of MLLMs in complex multimodal environments but also advances knowledge in the field.
        \begin{figure*}[t] %%
          \centering %?????????
          \includegraphics[width=1\textwidth]{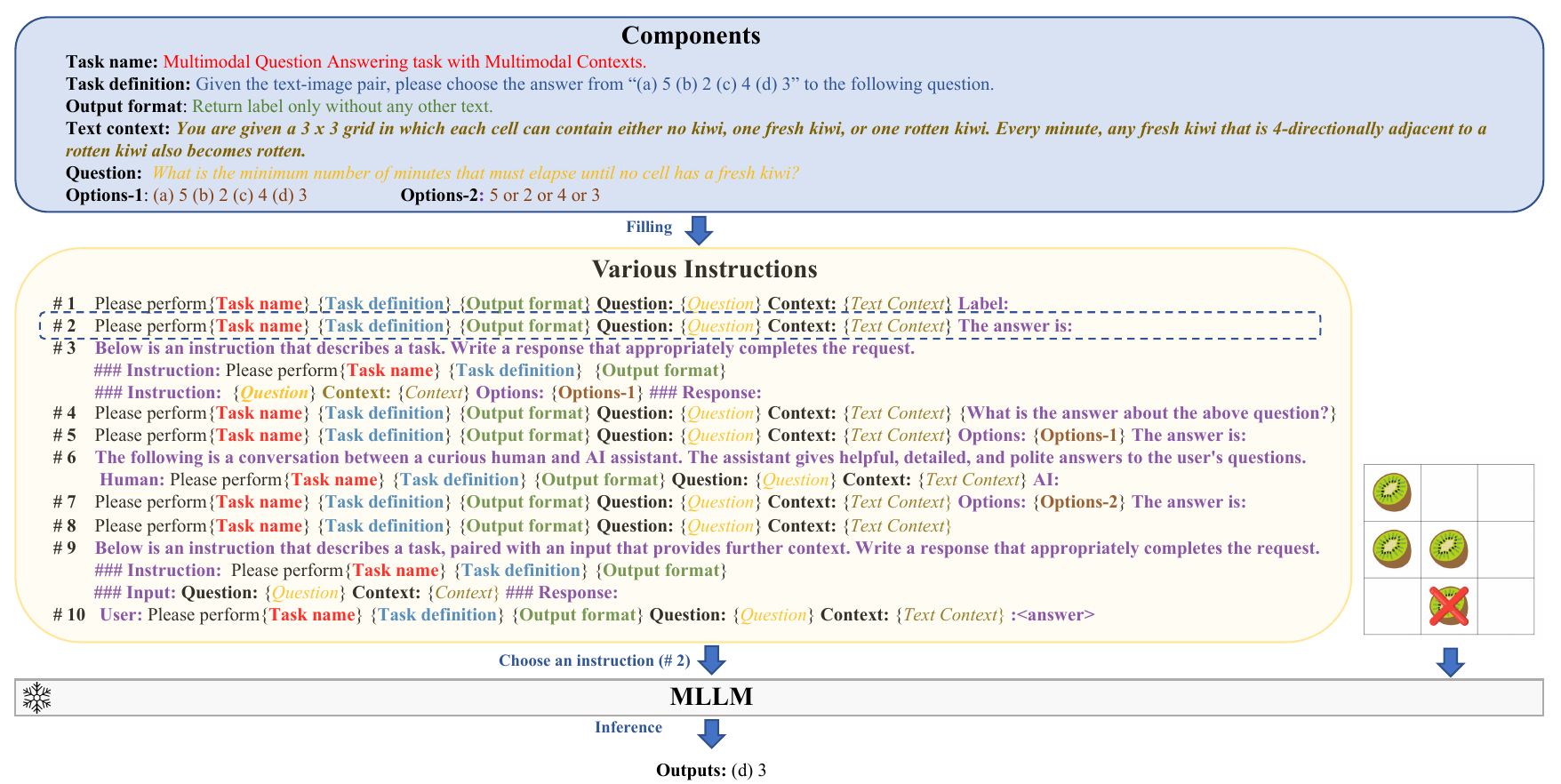}
          % \vspace{-1em}
          \caption{
        Inference process of Multimodal Language Models (MLLMs) for AlgoPuzzleVQA employing varied multimodal instructions. 
        We construct instructions based on these formats, encompassing mandatory components, such as \textbf{\textcolor{red}{Task name}}, \textbf{\textcolor[RGB]{70,130,180}{Task definition}}, and \textbf{\textcolor[RGB]{34,139,34}{Output format}}, \textbf{\textcolor[RGB]{255,215,0}{Question}},
          as well as optional components, for instance, \textbf{\textcolor[RGB]{128,128,0}{Context}} and \textbf{\textcolor[RGB]{205,133,63}{Options}}. 
          Furthermore, each format incorporates \textbf{\textcolor[RGB]{153,50,204}{Specific instruction trigger words}} customized for the respective instruction.
          Note that only text context are provided for inputting the Large Language Models (LLMs).
           }
          \label{Figure_1_PuzzleVQA_different_instructions} 
          %\vpsace{-1.5em}
        \end{figure*}

	\section{\textsc{MM-InstructEval}}
	\label{MM-InstructEval}
        We introduce \textsc{MM-InstructEval}, an extensive evaluation framework that integrates a diverse set of metrics to thoroughly assess various models and instructions in multimodal reasoning tasks involving vision-text contexts, as depicted in Figure \ref{Figure_MM-InstructEval-Framework}. In the following sections, we will delve into a detailed exploration of the definition of multimodal reasoning tasks that incorporate multimodal contexts. Additionally, we will outline the design of various instructions and discuss the metrics proposed for evaluating the performance of models and instructions.
        
        \subsection{Definition of Multimodal Reasoning Tasks with Vision-Text Contexts}
        Prior research on multimodal evaluation, highlighted in studies such as \citep{DBLP:journals/corr/abs-2306-13394, DBLP:journals/corr/abs-2307-16125, DBLP:journals/corr/abs-2306-09265, DBLP:journals/corr/abs-2308-02490}, often confines the evaluation of Multimodal Large Language Models (MLLMs) to a basic assessment of their multimodal capabilities. These studies typically examine the interplay between textual instructions/questions and vision-only contexts, primarily testing traditional vision-language (VL) capabilities \citep{DBLP:journals/corr/abs-2308-02490}, which encompass Language Generation, Recognition, Knowledge Reasoning, Spatial Reasoning, OCR, and more.
        A significant limitation of such evaluations is their inadequacy in fully assessing the capabilities of MLLMs to handle multimodal reasoning tasks that require combined information from both vision and text contexts. For example, in tasks like Multimodal Sarcasm Detection (MSD), shown in Figure \ref{Figure_multimodal reasoning Tasks}, detecting sarcasm through visual context alone is insufficient; accurate detection necessitates the integration of both vision and text contexts.
        It is crucial to recognize that tasks involving complex multimodal contexts demand substantial multimodal reasoning abilities. These tasks require extracting information from text, aligning it with visual data, and engaging in reasoning processes that inherently necessitate multimodal thinking.
        
\begin{figure*}[t] %%
          \centering %?????????
          \includegraphics[width=1\textwidth]{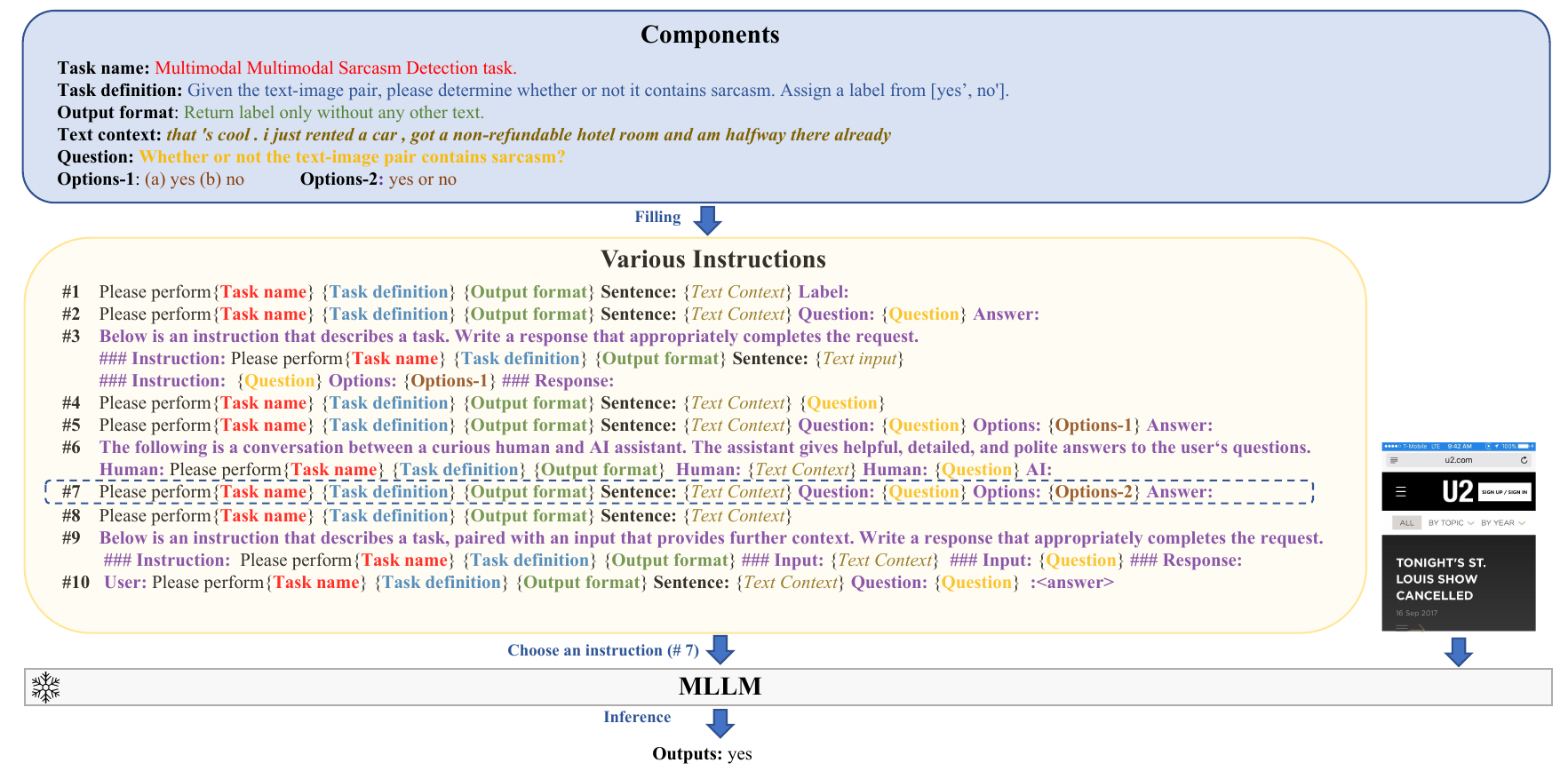} 
          %\vpsace{-1em}
          \caption{Inference process of Multimodal Language Models (MLLMs) for MSD employing varied multimodal instructions.
           }
          \label{Figure_1_MSD_Multimodal_Instruction} 
        \end{figure*}

        Inspired by MM-Vet \citep{DBLP:journals/corr/abs-2308-02490}, we delineate conventional vision-language (VL) capabilities for evaluation, with examples illustrated in the upper part of Figure \ref{Figure_multimodal reasoning Tasks}.
        In contrast, our research delves into multimodal reasoning tasks that incorporate vision-text contexts. These tasks require not just interactions between textual instructions and visual contexts to harness vision-language multimodal capabilities as described previously, but also an in-depth understanding of multimodal contexts, as demonstrated in Figure \ref{Figure_multimodal reasoning Tasks}. Illustrative examples are provided in Figure \ref{Figure_1_PuzzleVQA_different_instructions} and Figure \ref{Figure_1_MSD_Multimodal_Instruction}, showcasing how both textual content and images are crucial for formulating responses to the given instructions.
        We aim to conduct a comprehensive evaluation of various MLLMs, thoroughly covering a range of multimodal reasoning tasks that integrate vision-text contexts, encompassing Visual Question Answering with Multimodal Contexts (VQAMC), Multimodal Sentiment Analysis (MSA), Multimodal Aspect-Based Sentiment Analysis (MABSA), Multimodal Hateful Memes Detection (MHMD), Multimodal Sarcasm Detection (MSD), and Multimodal Relation Extraction (MRE).
        To our knowledge, it is the first work to evaluate such a diverse array of tasks using MLLMs. Notably, apart from VQAMC, existing benchmarks have not extensively explored the performance of MLLMs across various multimodal reasoning tasks possessing compelx multimodal contexts.
        % Since our work involves evaluating MLLMs, it inherently covers multimodal reasoning by leveraging interactions between textual instructions and multimodal contexts, similar to prior works such as \citep{DBLP:journals/corr/abs-2308-02490, DBLP:journals/corr/abs-2306-13394, DBLP:journals/corr/abs-2307-16125, DBLP:journals/corr/abs-2306-09265, DBLP:journals/corr/abs-2306-04757}.                   
        Ultimately, we delineate the essential Visual Language (VL) capabilities necessary for evaluating multimodal reasoning tasks with vision-text contexts, including Multimodal Alignment, Multimodal Fusion, Multimodal Content Understanding,  and Multimodal Reasoning. Illustrative examples of these capabilities are showcased in the lower part of Figure \ref{Figure_multimodal reasoning Tasks}.

        \begin{itemize} 
        \item[$\bullet$]{\textbf{Multimodal Alignment:} It refers to the process of associating or aligning information from different modalities in a coherent and meaningful manner. The objective is to establish correspondences or connections between different modalities to enable models to effectively leverage information from diverse sources. For instance, in multimodal models processing both images and text, alignment might involve matching specific words or phrases in the text with corresponding objects or concepts in the images. An example is aligning the phrase ``\textbf{Lady Gaga}'' with the ``\textbf{Person region}'' on the left side of the image in the ``MABSA'' column of Figure \ref{Figure_multimodal reasoning Tasks}. Such alignment fosters a more comprehensive and integrated understanding of the content, empowering the model to generate more contextually relevant responses or undertake tasks requiring a joint comprehension of multiple modalities.
        }
        \item[$\bullet$]{\textbf{Multimodal Fusion:} The process involves integrating information from multiple modalities to achieve a more comprehensive representation, thereby enhancing the overall understanding of multimodal data and improving model performance. For example, as depicted in Figure \ref{Figure_multimodal reasoning Tasks}, the fusion of the phrase ``\textbf{summer fun day}'' with the visual context of ``\textbf{the blue sunny sky}'' in the image within the ``MSA'' column amplifies the positive sentiment conveyed by the user. The fusion not only enriches the interpretation of individual modal inputs but also enables a more precise response from the model.
        }
        \item[$\bullet$]{\textbf{Multimodal Content Understanding:} 
         The capability involves the ability of a model to comprehend and interpret information presented across various modalities, such as the text and vision. For instance, in the ``MSD'' column of Figure \ref{Figure_multimodal reasoning Tasks}, there is a juxtaposition between the text context and the vision context, which collectively convey the irony of user. The scenario highlights the necessity for high-level understanding capabilities of the model to accurately interpret and respond to the complex, layered meanings embedded in multimodal inputs.
        }
        \item[$\bullet$]{\textbf{Multimodal Reasoning:} The process involves drawing logical conclusions or making predictions based on available information or evidence. In tasks that utilize multimodal contexts, the expected output may not be explicitly present in the original content. It requires the model to engage in reasoning based on the combined image and text contexts to derive the appropriate response. Such tasks demand inference from the integrated vision-text context to formulate the final answer, highlighting the need for sophisticated reasoning capabilities in handling complex multimodal data.
        }
        \end{itemize}

        \begin{figure*}[t] %%
          \centering %?????????
         
          \includegraphics[width=1\textwidth]{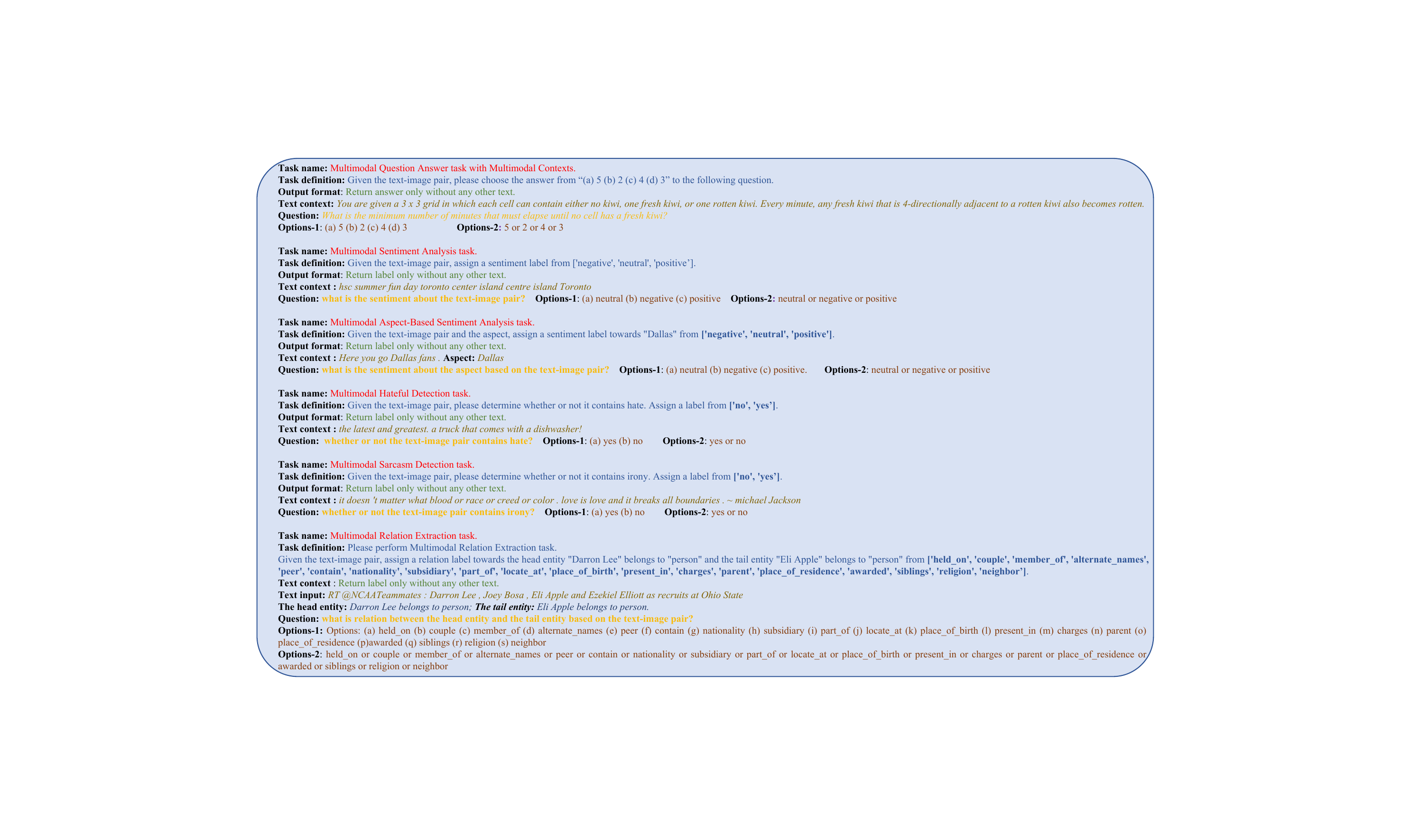} 
          %\vpsace{-1em}
           \caption{Details of the various components of multimodal instructions for different tasks, such as Multimodal Sentiment Analysis (MSA), Multimodal Aspect-Based Sentiment Analysis (MABSA), Multimodal Hateful Memes Detection (MHMD), Multimodal Sarcasm Detection (MSD), and Multimodal Relation Extraction (MRE).
           }
          \label{Figure_2_text_instruction_for_different_tasks} 
        \end{figure*}

        \subsection{Various Multimodal Instructions}
        Recent research in Natural Language Processing (NLP) has demonstrated that variations in instructions, even with identical semantics, can significantly affect the performance of a model \citep{DBLP:conf/nips/PerezKC21, DBLP:conf/acl/LuBM0S22, DBLP:journals/corr/abs-2305-15005, DBLP:journals/csur/LiuYFJHN23, gan2023sensitivity}.
        Building on the insights from \citeauthor{DBLP:journals/corr/abs-2305-15005}, we craft multimodal instructions for diverse multimodal reasoning tasks incorporating vision-text contexts, which include essential elements such as the task name, definition, and output format, as well as optional components like answer options. The text and image contexts form the multimodal content for each instance across various multimodal reasoning tasks.
        These instructions are specifically designed to evaluate how different MLLMs perform in a zero-shot setting when faced with diverse instructions. Although we develop only 10 manual instructions to gauge the performance of various models across different datasets, the evaluations offer valuable insights, enabling us to draw significant conclusions.
        For instance, we design a range of multimodal instructions for the Visual Question Answering with Multimodal Contexts task, as depicted in Figure \ref{Figure_1_PuzzleVQA_different_instructions}. Similarly, consistent instruction structures that integrate both text and image contexts also are created for other tasks. The instructions tailored for the Multimodal Sarcasm Detection (MSD) task are detailed in Figure \ref{Figure_1_MSD_Multimodal_Instruction}.
        
        Specifically, the \textbf{\textcolor{red}{Task name}} component serves to identify and clarify the purpose of each multimodal task. The \textbf{\textcolor[RGB]{70,130,180}{Task definition}} component, informed by the definition of task and annotation guidelines, outlines the label space as a set of options from which the model can generate responses. The \textbf{\textcolor[RGB]{34,139,34}{Output format}} component dictates the expected structure of the output of model, aiding in decoding the response into the desired format.
        The terms \textbf{\textcolor[RGB]{255,215,0}{Question}} and \textbf{\textcolor[RGB]{128,128,0}{Text context}} are essential in multimodal reasoning tasks involving vision-text contexts. The \textbf{\textcolor[RGB]{205,133,63}{Options}} component is optional and provides the model with multiple-choice questions, guiding it to respond accordingly.
        We observe that various MLLMs exhibit preferences for specific instructions. The preference may be influenced by the use of \textbf{\textcolor[RGB]{153,50,204}{Specific instruction trigger words}} during training of models, such as ``\#\#\# Instruction:'' or ``Question:''.
        Details of these components in text instructions for various tasks are illustrated in Figure \ref{Figure_2_text_instruction_for_different_tasks}.

        \subsection{Comprehensive Metrics}
        To effectively evaluate the performance of different models across various tasks, we develop a suite of comprehensive metrics. These metrics include the \textbf{Best Performance} metric, the \textbf{Mean Relative Gain} metric, the \textbf{Stability} metric, and the \textbf{Adaptability} metric. Each is designed to assess the efficacy of different models ($\mathcal{M}$) and various instructions ($\mathcal{I}$) across multiple datasets ($\mathcal{D}$), covering a range of diverse multimodal reasoning tasks with vison-text contexts.
        
        We choose Accuracy, denoted as \textbf{acc$_{mdi}$}, as primary metric to assess the performance of a model, $m \in \mathcal{M}$, with a specific multimodal instruction, $i \in \mathcal{I}$, on each dataset, $d \in \mathcal{D}$.
        %\vpsace{-1em}
        \begin{equation}
            \label{eq:acc}
            \begin{aligned}
                p^j_{mdi} = LM(T^j,V^j), \  acc_{mdi} = \frac{\sum_{j=1}^{N_d} {\mathbb{1}}_{(p^j_{mdi}=l^j_{mdi})}} {N_d},
            \end{aligned}
        \end{equation}
        where LM represents a specific language model, the LLM takes only the text instruction, including text context, as input, denoted as $T$, while the MLLM takes the multimodal context, $(T, V)$, as input. 
        % In this case, the visual context is optional for a particular LM. 
        For the $j$-th instance, $p^j_{mdi}$ represents the predicted label of the LM, $l^j_{mdi}$ is the true label. 
        $\mathbb{1}$ means the indicator function, $N_d$ is the number of instances, $acc_{mdi} \in Acc_{mdi}$ is the accuracy performance, $Acc_{mdi} \in \mathbb{R}^{|\mathcal{M}| \times |\mathcal{D}| \times |\mathcal{I}|}$, $|*|$ means the number of $*$. 
        $|\mathcal{M}|=35$, $|\mathcal{D}|=16$, and $|\mathcal{I}|=10$ represent the number of evaluated models, datasets, and instructions, respectively.
        %\vpsace{-0.5em}
        \paragraph{\textbf{Best Performance.}}
        To account for performance variations across different instructions, we report the\textbf{ best accuracy}, denoted as \textbf{$A^{\tilde{i}} \in \mathbb{R}^{|\mathcal{M}| \times |\mathcal{D}|}$}, which represents the highest accuracy achieved by each model among all instructions on each dataset. The metric highlights the upper performance limits for different models, providing insight into their optimal capabilities under various instructional scenarios. We acquire the best accuracy $A^{\tilde{i}}_{md}$. 
        
        %\vpsace{-0.5em}
        \begin{equation}
            \centering
            \begin{aligned}
            A_{md} & = Max(\{(acc_{mdi})\}_{i \in \mathcal{I}}), \\
             \tilde{i}_{md} & = \mathop{\arg\max}_{i}\ \ \ (\{(acc_{mdi})\}_{i \in \mathcal{I}}),
            \label{equ_best_acc}
            \end{aligned}
        \end{equation}
        %\vpsace{-0.5em}
        \paragraph{\textbf{Mean Relative Gain (MRG).}}
        % Given the variety of models and instructions, it is expected that we observe substantial variations in accuracy for each dataset, dependent on the specific models and instructions used.
        To effectively evaluate the overall performance across different models and instructions, aggregating metrics are introduced. Inspired by \citeauthor{DBLP:journals/corr/abs-2307-00259}, we employ two key measures:
        The  \textbf{Mean Relative Gains of Models} (\textbf{MRG$^{\mathcal{M}}$}) metric allows us to meaningfully compare and aggregate the performance of each model across all instructions. Specifically, \textbf{MRG$_{md}^{\mathcal{M}}$} quantifies the percentage by which the performance of model $m$ exceeds the average performance (across $\mathcal{M}$) on dataset $d$, when averaged across all instructions $\mathcal{I}$.
        It provides a clear metric to gauge how well each model performs relative to the average across all instructions for the specific dataset.
        % \vspace{-1.5em}
        \begin{equation}
          \begin{aligned}
                MRG^{\mathcal{M}}_{md}  = \frac{1}{|\mathcal{I}|} \sum_{i \in \mathcal{I}} r_{mdi}, MRG^\mathcal{M} \in \mathbb{R}^{|\mathcal{M}| \times |\mathcal{D}|}, \\
              r_{mdi}  = \frac{acc_{mdi} -\overline{Acc_{di}}}{\overline{Acc_{di}}} \times 100, \\
              \overline{Acc_{di}}  = \frac{1}{|\mathcal{M}|} \sum_{m \in \mathcal{M}} Acc_{mdi}, \overline{Acc_{di}} \in \mathbb{R}^{|\mathcal{D}| \times |\mathcal{I}|}. \\    
            \end{aligned}
            \label{Eq_model_MRG}
        \end{equation}
        The \textbf{Mean Relative Gains of Instructions} (\textbf{MRG$^{\mathcal{I}}$}) metric meaningfully compares and summarizes the performance of each instruction across all models, providing a clear view of which instructions generally yield better results regardless of the model used.
        \textbf{MRG}$_{id}^{\mathcal{I}}$ represents the percentage by which the performance of instruction $i$ on dataset $d$ exceeds the average performance, calculated across all models $\mathcal{M}$. The metric highlights the effectiveness of specific instructions relative to the norm, providing insights into which instructions stimulate model performance across diverse scenarios.
        %\vpsace{-0.5em}
        \begin{equation}
          \begin{aligned}
                MRG_{id}^{\mathcal{I}}  = \frac{1}{|\mathcal{M}|} \sum_{m \in \mathcal{M}} r_{idm}, MRG^\mathcal{I} \in \mathbb{R}^{|\mathcal{I}| \times |\mathcal{D}|}, \\
              r_{idm}  = \frac{acc_{idm} -\overline{Acc_{dm}}}{\overline{Acc_{dm}}} \times 100, \\
              \overline{Acc_{dm}}  = \frac{1}{|\mathcal{I}|} \sum_{i \in \mathcal{I}} Acc_{idm}, \overline{Acc_{dm}} \in \mathbb{R}^{|\mathcal{D}| \times |\mathcal{M}|}, \\
            \end{aligned}
            \label{Eq_Instruction_MRG}
        \end{equation}
        where the $Acc_{idm}$ is the transpose of $Acc_{mdi}$.
        \paragraph{\textbf{Stability.}}
        Stability is a crucial indicator of the reliability of both models and instructions. We assess the stability of a model across various instructions using the \textbf{Model Stability Metric}, denoted as \textbf{S$^{\mathcal{M^{'}}}$}. The metric is derived by calculating the standard deviation of the accuracy for each model when the instructions are varied for each dataset. A lower standard deviation indicates a model that performs consistently across different instructional contexts.
        \begin{equation}
            \begin{aligned}
                S^{\mathcal{M}^{'}}_{md} = \sqrt{\frac{1}{|\mathcal{I}|} \sum_{i \in \mathcal{I}} (acc_{mdi} - \overline{Acc_{md}})^2}.
            \end{aligned}
            \label{Eq_model_stability}
        \end{equation}
        Similarly, the stability of an instruction across different models is measured using the \textbf{Instruction Stability Metric}, represented as \textbf{S$^{\mathcal{I^{'}}}$}. The metric is calculated by computing the standard deviation of the accuracy for each instruction across different models for each dataset. Like S$^{\mathcal{M^{'}}}$, a lower standard deviation here signifies that the instruction performs consistently across various models.
        % Note that a few models may consistently perform poorly across all instructions, which results in good stability, albeit in a negative sense. Therefore, we evaluate the stability of LMs and instructions that perform well.
        
        \begin{equation}
            \begin{aligned}
                S^{\mathcal{I}^{'}}_{id} = \sqrt{\frac{1}{|\mathcal{M}|} \sum_{m \in \mathcal{M}} (acc_{idm} - \overline{Acc_{id}})^2},
            \end{aligned}
            \label{Eq_Instruction_stability}
        \end{equation}
        where $\overline{Acc_{md}}$ and $\overline{Acc_{id}}$ are the transpose of $\overline{Acc_{dm}}$ and $\overline{Acc_{di}}$, respectively. $\mathcal{M}^{'} \subset \mathcal{M}$, and $\mathcal{I}^{'} \subset \mathcal{I}$. 
        
        \paragraph{\textbf{Adaptability.}}
        Different instructions significantly influence model performance. To quantify the adaptability between models and instructions, we introduce the \textbf{Global Top-K Hit Ratio} (\textbf{$GHR@K$}) as a metric to evaluate how each model performs for various instructions across all datasets. The metric measures the frequency at which each instruction achieves Top-K performance on a specific model across all datasets. Our goal is to pinpoint specific instructions that enable each model to consistently achieve high performance across all datasets.
        Consequently, we assess each instruction based on its overall Top-K Hit Ratio score across all datasets for model $m$, providing a clear picture of which instructions are most effective universally.
        %\vpsace{-0.5em}
        \begin{equation}
            \begin{aligned}
             I^K_{d|m} = \mathop{\arg\max}_{i^1, ...  , i^K}K\ \ \ (\{(acc_{mdi})\}_{i \in \mathcal{I}}),  |I^K_{d|m}| = K, \\
            GHR@K_{m} = \frac{Counter_m^{\mathcal{I}}([I^K_{d^1|m} || ... || I^K_{d^{|\mathcal{D}|} |m}])}{(|\mathcal{D}| \times K)},
            \label{Eq_hit_ratio}
            \end{aligned}
        \end{equation}
        where $K=3$ represents the instructions needed to achieve the Top-K performance of model,
        $||$ is the contact operation, and
        ``Counter$_m^{\mathcal{I}}$'' is a dictionary function that counts the number of occurrences of each instruction $i \in \mathcal{I}$ for model $m$. $|GHR@K_{m}| = |\mathcal{I}|$.
        
        % \vspace{-1em}
        
        \section{Comprehensive Evaluation}
        In this section, we offer a comprehensive overview of the models evaluated by \textbf{\textsc{MM-InstructEval}}, as well as the various multimodal reasoning tasks with vision-text contexts and their corresponding datasets. Specifically, we assess a range of language models (LMs) across different multimodal reasoning tasks, as detailed in Table \ref{Table_dataset_statistic}. It includes both pure Large Language Models (LLMs) and Multimodal Large Language Models (MLLMs). We consider both closed and open-source models within each category, aiming to include as many currently popular models as possible. Detailed information about these models is provided in Table \ref{Table_different_models}.
        \textbf{\textsc{MM-InstructEval}} is designed to be flexible, easily accommodating emerging models and extendable to new datasets. We continuously update and enrich it. Detailed and latest results can be found on the \href{https://declare-lab.github.io/MM-InstructEval/}{\textsc{MM-InstructEval} Leaderboard}.
        Our evaluations are performed in a zero-shot setting, aiming to assess the capabilities of models in multimodal reasoning tasks with multimodal contexts, enabling them to reason accurate answers without the need for fine-tuning or few-shot demonstrations on our benchmarks. All experiments\footnote{Different seeds have a limited impact on the experimental results. We only use \textbf{seed 42} for experiments, which allows us to still draw valuable conclusions.} are performed using NVIDIA A6000 GPUs.
        
        \subsection{Why do Zero-Shot Evaluation?}
        Following previous research \citep{DBLP:journals/corr/abs-2306-13394, liu2023mmbench, DBLP:journals/corr/abs-2307-16125, DBLP:journals/corr/abs-2306-09265, DBLP:journals/corr/abs-2308-02490, yue2023mmmu, ge2023mllm} that evaluates MLLMs, we concentrate on assessing zero-shot performance across a variety of multimodal reasoning tasks with vision-text contexts for the following reasons:
        \begin{itemize}
        \item Zero-shot evaluation serves as a fundamental and rigorous test, measuring the intrinsic ability of models to handle diverse multimodal reasoning tasks with complex contexts. It allows us to draw valuable conclusions about the capabilities of models without prior specialized training.
        \item Most MLLMs are designed to process single image-text pairs, and the format of ``multiple image-text pairs'' does not align with their training configurations, indicating limited compatibility with few-shot evaluation settings.
        \item Preliminary experiments to assess the in-context capabilities of a few MLLMs yielded suboptimal results. Looking ahead, we plan to develop a tailored subset of data and design effective methods to facilitate in-context evaluation, aiming to enhance the performance of MLLMs in such scenarios.
        \end{itemize}

        \subsection{Evaluated Models}
        Below is a detailed overview of the various language models employed in our evaluation.
        \begin{table*}[t] \small
        \begin{center}
        \renewcommand{\arraystretch}{1} 
        \caption{
        A comprehensive summary of various models (\textbf{\textcolor{red}{Red}} indicates closed-source models; others are open-source). Below are the abbreviations used in the table:
        `\textbf{LLMs}' refers to the pretrained Language Models (LLMs) that form the backbone of the Multimodal Large Language Models (MLLMs).
        `\textbf{PVM}' signifies the pretrained visual model backbone of the MLLMs.
        `\textbf{To-Paras}' and `\textbf{Tr-Paras}' represent the total number of parameters and trainable parameters for each language model, respectively.
        `\textbf{Held-In}' refers to the dataset that each specific MLLM was trained or fine-tuned on.
        The `\textbf{GPU}' column indicates the single GPU utilization during inference, and the `\textbf{Time}' column signifies the time taken for model inference on each text/multimodal instance using a single GPU. Note that GPU usage and inference time may vary slightly across different datasets due to varying data lengths. For these two metrics, we provide an approximate mean value across all datasets for each model.
        A dash, `-', indicates that the data is not applicable or not involved, and `Unknown' means we do not have information on whether these models cover the evaluated dataset.
        }
        %\vpsace{-1em}
        \resizebox{1\linewidth}{!}{
        \begin{tabular}{
        p{2.1cm}< \centering| 
        p{3.7cm}< \centering| p{4cm}< \centering p{3.5cm}< \centering
        p{1.4cm}< \centering p{1.4cm}< \centering  p{1.4cm}< \centering 
        p{0.6cm}< \centering p{0.6cm}< \centering}
        \toprule[1pt]
        \textbf{Modality} &
        \textbf{Models} & \textbf{LLMs} & \textbf{PVMs} &
        \textbf{To-Paras} & \textbf{Tr-Paras} & \textbf{Held-In}
        & \textbf{GPU} & \textbf{Time}\\
        % \cline{1-5}
        \midrule[1pt]
        \multirow{7}{*}{\textbf{Text}}
        & \textbf{\textcolor{red}{ChatGPT}} & gpt-3.5-turb &	- & - & - & Unknown & -& - \\
        & \textbf{LLaMA1-7B}& LLaMA-V1-7B & - & 6.74B & 6.74B & - & 26G & 2.0s \\
        & \textbf{LLaMA1-13B} & LLaMA-V1-13B & - & 13.02B & 13.02B & - & 48G & 9.0s\\
        & \textbf{LLaMA2-7B} & LLaMA-V2-7B & - & 6.74B & 6.74B  & - &26G & 1.0s \\
        & \textbf{LLaMA2-13B} & LLaMA-V2-13B & - & 13.02B & 13.02B& - & 48G & 8.0s\\
        & \textbf{LLaMA3.1-8B-Instruct} & LLaMA3.1-8B & - &  8.03B & 8.03B & Unknown & 32G & 2.5s   \\ 
        & \textbf{Mixtral-AWQ} & Mixtral-8x7B-Instruct-v0.1 & - &  0.26B &  0.26B & Unknown & 25G & 5s \\
        & \textbf{Gemma} & Gemma-7B-it & - & 8.54B & 8.54B & Unknown & 36G & 1.5s \\
        & \textbf{Flan-T5-XXL}	& Flan-T5-XXL & - & 11.14B & 11.14B & - & 44G & 0.3s  \\
        \midrule[1pt]
        \multirow{5}{*}{\textbf{Closed MLLMs}}
        & \textbf{\textcolor{red}{GPT-4V$\heartsuit$}} & gpt-4 & vision-preview  & - & - & Unknown & - & -\\
        & \textbf{\textcolor{red}{GPT-4o$\heartsuit$}} & gpt-4o & - & - & - & Unknown & - & - \\
        & \textbf{\textcolor{red}{Claude3$\heartsuit$}} & claude-3-opus-2024022 & - & - & - & Unknown & - & -\\
        & \textbf{\textcolor{red}{GLM-4V-plus$\heartsuit$}} & GLM-4 & - & - & - & Unknown & - & - \\
        & \textbf{\textcolor{red}{Gemini-V}} & gemini-pro-vision & - & - & - & Unknown & - & -\\
        \midrule[1pt]
        \multirow{30}{*}{\textbf{Open  MLLMs}}
        & \textbf{OpenFlamingo} &  LLaMA-7B & ViT-L/14 & 8.34B & 1.31B & & 34G & 1.5s\\
        & \textbf{Fromage} & OPT-6.7B & ViT-L/14 & 6.97B & 0.21B & - &14G  & 5.0s\\
        & \textbf{LLaVA-v0-7B} & Vicuna-7B-v0 & ViT-L/14 & 6.74B	& 6.74B	& - & 15G & 2.5s\\
        & \textbf{LLaVA-v0-13B} & Vicuna-13B-v0 & ViT-L/14 & 13.02B & 13.02B & - & 27G & 2.0s\\ 
        & \textbf{LLaVA-v1.6-7B} & Vicuna-7B & ViT-L/14  &  7.06B & 6.76B & -&  26G & 1.3s \\
        & \textbf{LLaVA-v1.6-13B} & Vicuna-13B & ViT-L/14 &  13.35B & 13.05B &- &  36G & 1.3s \\
        & \textbf{MiniGPT4} & Vicuna-13B &  ViT-g/14 & 14.11B & 0.04B & - & 15G & 1.3s\\
        & \textbf{mPLUG-Owl} & LLaMA-V1-7B & ViT-L/14 & 7.12B & 7.12B	& - &16G & 1.0s \\
        & \textbf{mPLUG-Owl2.1} & Qwen-7B &  ViT-G/14 &  11.42B & 11.42B & - & 24G & 1.0s\\
        & \textbf{LLaMA-Adapter V2} & LLaMA-V1-7B & ViT-L/14 & 7.23B & 7.23B & - & 14G & 1.3s\\
        & \textbf{VPGTrans} & Vicuna-7B & -  & 7.83B &	0.11B & - & 36G & 10s\\
        % & \textbf{Otter} &  LLaMA-7B & ViT-L/14  & - & -& 40G & 2.5s\\
        & \textbf{Multimodal-GPT} &  LLaMA-V1-7B & ViT-L-14 & 8.37B & 0.02B & - &	18G	& 0.5s \\
        & \textbf{LaVIN-7B} & LLaMA-V1-7B & ViT-L/14 & 7.17B &	7.17B & ScienceQA & 16G & 4.0s  \\
        & \textbf{LaVIN-13B} &  LLaMA-V1-13B & ViT-L/14 & 13.36B & 13.36B	& ScienceQA & 28G& 11.0s \\
        & \textbf{Lynx} & Vicuna-7B & Eva-ViT-1b & 8.41B & 0.69B & Hate & 44G & 6.5s \\
        & \textbf{Fuyu-8B} & Persimmon-8B  & - & 9.41B & 9.41B & - & 44G & 1.3s \\
        & \textbf{LaVIT} &  LLaMA-V2-7B & ViT-G/14 & 8.30B & 0.41B & - & 23G & 2.2s \\
        & \textbf{LLaMA-3.2-11B-Vision} & LLaMA-3.1 & Vision Adapter & 10.67B & 10.67B & -  & 23G & 3s \\
        & \textbf{MiniCPM-V2.6} & Qwen2-7B & SigLip-400M & 8.10B & 8.10B & ScienceQA & 17G & 0.7s \\
        % \midrule
        % \multirow{2}{*}{\textbf{Multimodal}}
        & \textbf{BLIP-2} & Flan-T5-XXL & ViT-G/14 & 12.23B & 0.11B	&  - & 26G & 3.5s\\
        & \textbf{InstructBLIP} & Flan-T5-XXL & ViT-G/14 & 12.31B & 0.45B & - & 16G	& 0.3s\\
        & \textbf{GLM-4V-9B} & GLM-4-9B &  EVA-CLIP-E & 13.91B & 13.91B & ScienceQA & 30G & 2.4s \\
        & \textbf{InternVL2.5-8B} & InternViT-300M-448px-V2\_5 & internlm2\_5-7b-chat & 8.08B & 8.08B & ScienceQA & 24G & 2s\\
        & \textbf{DeepSeek-VL2-tiny} & MoE-based LLM & SigLIP-SO400M& 3B & 0.57B & - & 20G & 0.2s \\
        & \textbf{DeepSeek-VL2-small} & MoE-based LLM & SigLIP-SO400M& 16B & 2.4B & - &  48G & 0.25s\\
        & \textbf{DeepSeek-VL2} & MoE-based LLM & SigLIP-SO400M & 27B & 4.1B  & - & - & -\\
        & \textbf{Qwen-VL-Chat} & Qwen-7B & ViT-bigG & 9.66B & 9.66B & - & 22G & 1.0s \\
        & \textbf{Qwen2-VL-7B-Instruct} &  Qwen2-7B & ViT-G/14 & 8.29B & 8.29B & - & 18G & 0.9s \\
        & \textbf{Qwen2-VL-72B-Instruct} & Qwen2-72B & ViT-G/14 & - & - & - & - & -\\
        & \textbf{Qwen2.5-VL-3B-Instruct} & Qwen2.5-3B & ViT & 3.75B & 3.75B & - & 10G & 0.25s \\
        & \textbf{Qwen2.5-VL-7B-Instruct} & Qwen2.5-7B & ViT & 8.29B & 8.29B & - & 20G & 0.5s \\
        \bottomrule[1pt]
        \end{tabular}}
        
        \label{Table_different_models}
        \end{center}
        % \vspace{-1em}
        \end{table*}

        \subsubsection{LLMs}
        We evaluate eight LLMs (\textcolor{red}{closed-source} and open-source) across multiple tasks using text-only contexts. Since the questions from the AlgoPuzzleVQA and MMMU datasets require visual content for answers, we do not include these two datasets in our evaluation of LLMs.
        \begin{itemize}
            \item {
            \textbf{\textcolor{red}{ChatGPT}} \citep{chatgpt} is a closed-source conversational AI language model developed by OpenAI, renowned for its impressive performance across a wide range of NLP tasks. Specifically, we evaluate the classic model, ``ChatGPT (gpt-3.5-turbo)''\footnote{Our evaluation of ChatGPT is carried out between July and September 2023. }.}
            \item {
            The LLaMA family of models includes LLaMA-1 \citep{DBLP:journals/corr/abs-2302-13971}, LLaMA-2 \citep{DBLP:journals/corr/abs-2307-09288}, and LLaMA-3.1. We assess the performance of the ``decapoda-llama-7b/13b-hf'', ``meta-Llama-2-7b/13b-hf'', and ``meta-llama-Llama-3.1-8B-Instruct'' models.}
            \item{
            \textbf{Mixtral} \citep{Mixtral} is a sparse mixture-of-experts network (SMoE). Due to its high GPU resource requirements, which exceed our server capabilities, we employ a version with weight quantization using AWQ\footnote{AWQ is an efficient, accurate, and extremely fast low-bit weight quantization method that supports 4-bit quantization.}. Specifically, we evaluate the ``ybelkada/Mixtral-8x7B-Instruct-v0.1-AWQ'' model in our study. 
         }
            \item{
            \textbf{Gemma} \citep{Gemma} represents a new family of state-of-the-art open-source LLMs by Google, developed under the Gemini initiative\footnote{\url{https://deepmind.google/technologies/gemini/\#gemini-1.5}}. In our study, we evaluate the ``google/gemma-7b-it'' model.
            }
            \item{
            \textbf{Flan-T5} \citep{DBLP:journals/corr/abs-2210-11416} is a model that extends its capabilities to 1836 fine-tuning tasks through instruction tuning, enhancing the performance and usability of the model. In our evaluation, we specifically examine the ``flan-t5-xxl'' version.}
        \end{itemize}

        \subsubsection{MLLMs}
        \label{subsubsection:MLLLMs}
        We evaluate a total of 36 Multimodal Large Language Models. Among them, five \textcolor{red}{closed-source} models are included, such as GPT-4V \citep{GPT4V}, Gemini-V \citep{Gemini}, and Claude3 \citep{Claude}. Given that GPT-4V is costly and has limited access, we restrict our evaluation to a subset of the test dataset.
        The subset, marked with ${\heartsuit}$, is sampled by the Consistently Distributed Sampling (CDS) approach, ensuring a distribution consistent with the original \citep{yang2023few_MultiPoint, yang2023few_GMP}. CDS ensures that the class distribution of the few-shot dataset closely aligns with that of the full dataset. This method creates a representative sub-dataset that accurately reflects the true distribution of the original dataset. Specifically, we randomly select approximately 10\% of the test dataset as a subset, preserving the class distribution of the full training dataset. For instance, in a sentiment analysis dataset, the ratio of positive, neutral, and negative sentiment data is maintained at 4:1:2.

        \begin{itemize}
            \item{
            \textbf{\textcolor{red}{GPT-4V$^{\heartsuit}$}} \citep{GPT4V} extends ChatGPT's capabilities to analyze image inputs, enhancing its functionality. We evaluate the ``gpt-4-vision-preview''\footnote{Our evaluation of GPT-4V via the corresponding API is carried out in February 2024.} version. }

            \item{
            \textbf{\textcolor{red}{GPT-4o$^{\heartsuit}$}} \citep{hurst2024gpt} is an autoregressive omni-model trained end-to-end across text, vision, and audio. This unified architecture processes all inputs and outputs through the same neural network, enabling seamless multimodal understanding. The ``gpt-4o\footnote{Our evaluation of GPT-4o via the corresponding API is carried out in February 2025.}'' is evaluated.}
            
            \item {\textbf{\textcolor{red}{Claude3-V$^{\heartsuit}$}} represents the latest in cutting-edge closed-source models. Claude \citep{Claude}, developed by Anthropic, is a family of large language models designed to transform interactions with AI. For our evaluation, we utilize the ``claude-3-opus-20240229''\footnote{Our evaluation of Claude3-V via the corresponding API is carried out in March 2024.} version.
            }
            \item{
            \textbf{\textcolor{red}{Gemini-V}} is capable of processing both text and image inputs to generate text responses. Gemini \citep{Gemini}, developed by Google, is a series of advanced multimodal generative AI models. In our study, we evaluate the ``gemini-pro-vision''\footnote{Our evaluation of Gemini-V via the corresponding API was conducted in January 2024.} version.
            }

            \end{itemize}

        The open-source MLLMs (31 models) are developed by integrating pretrained visual models (PVMs) with LLM backbones.
        Typically, the PVMs are derived from pretrained CLIP models, such as ViT-L/14, ViT-g/14, etc. The LLMs span various families including the LLaMA family, with models like LLaMA-V1 and LLaMA-V2, Vicuna, the Flan-T5 family, and others.
        \begin{itemize}
            \item { 
            The \textbf{OpenFlamingo} family of models is renowned for its impressive performance in few-shot learning across various open-ended vision and language tasks, as detailed in \citep{alayrac2022flamingo, Awadalla2023openflamingo}. We evaluate the pre-trained ``Openflamingo-9B'' model.
            }
            \item{
            \textbf{Fromage} is trained by visually grounding LLMs through image captioning and contrastive learning \citep{koh2023grounding}. We assess the pre-trained  ``fromage-model''.
            }
            \item{
            \textbf{LLaVA} \citep{DBLP:journals/corr/abs-2304-08485} is an end-to-end traine MLLM that connects a vision encoder with an LLM for extensive visual and language understanding.
            We assess different pretrained LLaVA models, which are built on various scales of LLMs, including ``LLaVA-v0-7B'', ``LLaVA-v0-13B'', ``LLaVA-v1.6-7B'', and ``LLaVA-v1.6-13B''.
            }
            \item{
             \textbf{MiniGPT-4} aligns a frozen visual encoder with a frozen LLM called Vicuna, using one projection layer \citep{DBLP:journals/corr/abs-2304-10592}. We utilize the pretrained ``MiniGPT-4 checkpoint with Vicuna 13B''.
            }
            \item{
            \textbf{mPLUG-Owl} equips LLMs with multi-modal abilities through modularized learning of a foundation LLM, a visual knowledge module, and a visual abstractor module \citep{DBLP:journals/corr/abs-2304-14178, ye2023mplug}. We evaluate the ``mPLUG-Owl 7B'' and ``mPLUG-Qwl2.1''.
            }
            \item{
            \textbf{LLaMA-Adapter V2} \citep{DBLP:journals/corr/abs-2304-15010} is jointly trained on image-text pairs and instruction-following data. We evaluate the ``LLaMA-Adapter V2 Multimodal'' model.
            }
            \item{
            \textbf{VPGTrans} \citep{DBLP:journals/corr/abs-2305-01278} is a two-stage transfer framework facilitates the efficient transfer of Visual Prompt Generators (VPG) across LLMs while minimizing the need for extensive training data. We assess the performance of the ``VL-LLaMA'' model, in which the VPG is seamlessly transferred from BLIP-2 OPT6.7B to LLaMA-7B.
            }
            \item{
            \textbf{Multimodal-GPT} \citep{DBLP:journals/corr/abs-2305-04790} incorporates the Low-rank Adapter (LoRA) \citep{DBLP:conf/iclr/HuSWALWWC22} in both the gated-cross-attention and self-attention components of the language model. We assess the pretrained ``mmgpt-lora-v0-release'' weights.
            }
            \item{\textbf{LaVIN} \citep{DBLP:journals/corr/abs-2305-15023} is a model proposed based on the concept of Mixture-of-Modality Adaptation (MMA). We conduct evaluation of the ``LaVIN-7B'' and ``LaVIN-13B''.}
            \item{
            \textbf{Lynx} \citep{DBLP:journals/corr/abs-2307-02469} is a straightforward prefix-tuning GPT4-style model featuring a two-stage training approach. We utilize the ``finetune-lynx.pt'' model for evaluation.
            }
            \item{
            \textbf{Fuyu-8B} \citep{Fuyu} is a multimodal text and image converter trained by Adept AI\footnote{\url{https://www.adept.ai/}}. Architecturally, Fuyu is a pure decoder transformer, which is no image encoder and instead utilizes a transformer decoder. ``adept/Fuyu-8B'' is assessed.
            }
            \item{
            \textbf{LaVIT} \citep{jin2023unified} is a new effective, general-purpose multimodal foundation model that goes beyond traditional adapter-based architectures. ``rain1011/LaVIT-7B-v2'' is evaluated in our paper.
            }
            \item{
            \textbf{LLaMA 3.2-Vision} \citep{Llama-3_2-11B-Vision} comprises a collection of multimodal language models pretrained and instruction-tuned for visual tasks, including image recognition, reasoning, captioning, and visual question answering. In our evaluation, we focus on the ``LLaMA-3.2-11B-Vision'' model.
            }
            \item {
            \textbf{MiniCPM-V} \citep{yao2024minicpm} is a series of efficient multimodal language models designed for edge devices, striking a balance between performance and computational efficiency. We evaluate ``MiniCPM-V2.6'', the latest and most capable model in this series.
            }
            \item{
            \textbf{BLIP-2} \citep{DBLP:journals/corr/abs-2301-12597} addresses the modality gap through the use of a lightweight Querying Transformer that connects a frozen pre-trained image model with a language model. We consider the version of this model labeled as ``blip2-flan-t5-xxl''.
            }
            \item{
            \textbf{InstructBLIP} \citep{dai2023instructblip} conducts a comprehensive and systematic study on vision-language instruction tuning by utilizing pretrained BLIP-2 models.  We utilize the pretrained model ``blip2-instruct-Flan-T5xxl''.
            }
            \item{\textbf{GLM-4V} \citep{hong2024cogvlm2} is a series of multimodal large language models (MLLMs) developed by Zhipu AI. These bilingual visual-language models are designed to explore image understanding capabilities in both English and Chinese. In our evaluation, we assess ``THUDM/glm-4v-9b'' and ``\textbf{\textcolor{red}{glm-4v-plus$^{\heartsuit}$}}''. Notably, glm-4v-plus\footnote{Our evaluation of glm-4v-plus via the corresponding API is carried out in February 2025.} is evaluated on a subset of data due to its high computational cost.}
            
            \item{\textbf{InternVL2.5-8B}} \citep{chen2024internvl}  retains the same model architecture as its predecessors, InternVL 1.5 and 2.0, following the ``ViT-MLP-LLM'' paradigm. The ``OpenGVLab/InternVL2.5-8B'' is assessed.
            \item{\textbf{DeepSeek-VL2}} \citep{wu2024deepseekvl2mixtureofexpertsvisionlanguagemodels} is an advanced series of large Mixture-of-Experts (MoE) Vision-Language Models that significantly improves upon its predecessor. In this evaluation, the models ``deepseek-ai/deepseek-vl2-tiny'', ``deepseek-ai/deepseek-vl2-small'', and ``deepseek-ai/deepseek-vl2'' are assessed. Note that we use the API for the deepseek-vl2 model due to limited computing resources.
            \item{
            \textbf{Qwen-VL Family} \citep{bai2023qwen} represents a series of highly performant and versatile vision-language foundation models derived from the Qwen series \citep{Qwen}. This family introduces new visual receptors, including Qwen-VL, Qwen2-VL \citep{Qwen2VL}, and the latest iteration, Qwen2.5-VL \citep{Qwen2.5-VL}. In our evaluation, we focus on various model variants, i.e., ``Qwen-VL-Chat'' and ``Qwen2-VL-7B-Instruct'', ``Qwen2-VL-72B-Instruct'', ``Qwen2.5-VL-3B-Instruct'', and ``Qwen2.5-VL-7B-Instruct''. We also use the API for the Qwen2-VL-72B-Instruct model due to limited computing resources.
            }
            
        \end{itemize}

        \subsection{Multimodal Datasets}
        \begin{table}[t] 
        \begin{center}
        \renewcommand{\arraystretch}{1.1} 
        \caption{Evaluation datasets statistics for different tasks. ``Labels'' denotes the number of labels among each task. ``-'' means that the VQAMC task has no fixed label space, and each question has a unique answer.
        }
        \resizebox{1\linewidth}{!}{
        \begin{tabular}{
        p{1cm}< \centering| p{2.8cm}< \centering p{1.8cm}< \centering 
        p{0.5cm}< \centering p{0.7cm}< \centering p{0.6cm}< \centering }
        \toprule[1pt]
        \textbf{Task} & \textbf{Dataset} & \textbf{Modality} & \textbf{Test} & \textbf{Labels} & \textbf{Metric}\\
        % \cline{1-5}
        \midrule[1pt]
        \multirow{3}{*}{\textbf{VQAMC}}
         & ScienceQA & Text-Image & 2017 & -  & Acc \\
         & AlgoPuzzleVQA & Text-Image & 1800 & -  & Acc \\
         & MMMU-val & Text-Image & 900 & -  & Acc \\
        \midrule[0.5pt]
        \multirow{7}{*}{\textbf{MSA}}
        & MVSA-Single  & Text-Image &413 & 3 & Acc \\
        & MVSA-Multiple  & Text-Image & 1531 & 3 & Acc \\
        & TumEmo  & Text-Image & 9463  & 7 & Acc \\
        & MOSI-2  & Video & 654 & 2 & Acc\\
        & MOSI-7  & Video & 684 & 7 & Acc \\
        & MOSEI-2  & Video & 2797 & 2 & Acc\\
        & MOSEI-7  & Video & 3588 & 7 & Acc \\
        \midrule[0.5pt]
        \multirow{3}{*}{\textbf{MABSA}}
        & Twitter-2015  & Text-Image & 1037 & 3 & Acc \\
        & Twitter-2017  & Text-Image & 1234 & 3 & Acc \\
        & MASAD  & Text-Image & 4935 & 2 & Acc \\
        \midrule[0.5pt]
        \textbf{MHMD}  & Hate  & Text-Image & 500 & 2 & Acc \\
        \midrule[0.5pt]
        \textbf{MSD} & Sarcasm   & Text-Image & 2409 & 2 & Acc \\
        \midrule[0.5pt]
        \textbf{MRE}  & MNRE   & Text-Image & 640 & 19 & Acc \\
        \bottomrule[1pt]
        \end{tabular}
        }
        \label{Table_dataset_statistic}
        \end{center}
        % \vspace{-1em}
        \end{table}

        We conduct comprehensive evaluation of various Language Models across a diverse range of multimodal reasoning tasks, including \textbf{VQAMC}, \textbf{MSA}, \textbf{MABSA}, \textbf{MHMD}, \textbf{MSD}, and \textbf{MRE}. Detailed statistics for each task and the datasets can be found in Table \ref{Table_dataset_statistic}. The detailed introduction to various multimodal reasoning tasks with vision-text contexts and related datasets are as follows.
        \begin{itemize}
           \item{\textbf{Visual Question Answering with Multimodal Contexts (VQAMC)}:
           \textbf{ScienceQA} \citep{DBLP:conf/nips/LuMX0CZTCK22} is a well-known visual question-answering dataset featuring a variety of science topics. It provides both image-text contexts, which can range from semantically rich information to simple hints. Given our evaluation, we specifically utilize the portion of the dataset that includes image context.
           \textbf{AlgoPuzzleVQA} \citep{chia-etal-2024-puzzlevqa} is a novel dataset designed to challenge and evaluate the capabilities of MLLMs in solving algorithmic puzzles that require visual understanding, language comprehension, and complex algorithmic reasoning. The dataset includes areas such as boolean logic, combinatorics, graph theory, optimization, and search. It is crucial to note that the AlgoPuzzleVQA dataset requires the integration of textual context with visual elements to resolve specific queries effectively. Therefore, our assessment focuses exclusively on MLLMs capable of handling such intricate, dual-modality data interpretation. For simplicity, we refer to this dataset as ``PuzzleVQA''.    
           \textbf{MMMU} \citep{yue2023mmmu} presents significant challenges as an extensive benchmark designed for college-level multi-discipline multimodal understanding and reasoning. It encompasses six common disciplines: Art \& Design, Business, Science, Health \& Medicine, Humanities \& Social Science, and Tech \& Engineering. Our evaluations specifically focus on the validation set of MMMU, succinctly referred to as ``MMMU-val''.
           }
            \item{\textbf{Multimodal Sentiment Analysis (MSA)}:
            MSA aims to detect the overall sentiment of a text-image pair or a video, addressing the complexity of sentiments expressed across different modalities \citep{DBLP:conf/sigir/XuMC18, DBLP:journals/tmm/YangFW021, DBLP:conf/acl/YangF0W20, DBLP:conf/naacl/LiXZZ22, Yang2023MultipleCL}.
            Our evaluation includes three widely used text-image datasets: MVSA-Single and MVSA-Multiple \citep{DBLP:conf/mmm/NiuZPE16}, and TumEmo \citep{DBLP:journals/tmm/YangFW021}.
            Additionally, we assess four video datasets: MOSI-2, MOSI-7 \citep{DBLP:journals/corr/ZadehZPM16}, MOSEI-2, and MOSEI-7 \citep{DBLP:conf/acl/MorencyCPLZ18}.
            Given that most MLLMs primarily accept text-image pairs, for video datasets, we initially extract one frame per second to form a set of candidate frames. We then randomly select a single frame\footnote{We also experiment with selecting multiple frames, such as three, and inputting them individually into the MLLMs, using a voting principle to determine the final result. However, results are comparable to using a single randomly chosen frame. Thus, we opt for one frame as the visual input for each video.} to serve as the image input for MLLMs.
            For MOSI-2 and MOSEI-2, the label space includes \{positive, negative\}, excluding neutral, which is labeled as zero. In contrast, MOSI-7 and MOSEI-7 feature a label space comprising \{strongly positive, positive, weakly positive, neutral, weakly negative, negative, strongly negative\}, with neutral sentiment explicitly recognized.
            }
            \item{\textbf{Multimodal Aspect-Based Sentiment Analysis  (MABSA)}:
             MABSA focuses on identifying sentiments associated with specific aspect terms within the given vision-text contexts \citep{DBLP:conf/acl/HuPHLL19, DBLP:conf/ijcai/Yu019, DBLP:conf/emnlp/JuZXLLZZ21, DBLP:conf/acl/LingYX22, DBLP:conf/emnlp/YangZ022, DBLP:journals/taffco/YuCX23, DBLP:journals/ipm/YangNY22, DBLP:conf/mm/YuZL22}. Our experiments utilize three widely recognized datasets:
            Twitter-2015 \citep{DBLP:conf/aaai/0001FLH18}, 
            Twitter-2017 \citep{DBLP:conf/acl/JiZCLN18}, and 
            MASAD \citep{DBLP:journals/ijon/ZhouZHHH21}.
            These datasets evaluate the effectiveness of MLLMs in accurately discerning sentiments related to specific aspects based on diverse multimodal contexts.}
            \item{\textbf{Multimodal Hateful Memes Detection (MHMD)}:
            A new challenge set for multimodal classification, specifically designed for MHMD, is introduced in \citep{mathias2021findings}. However, a labeled test set has not been publicly released for this challenge. Consequently, our evaluation focuses on the publicly available labeled validation set, commonly referred to as ``dev-seen'' in the literature. For simplicity, we refer to the dataset as \textbf{Hate} in this paper.
            }
            \item{\textbf{Multimodal Sarcasm Detection (MSD)}:
            MSD aims to identify sarcasm with multimodal contents, drawing upon the interplay of text and visual cues \citep{DBLP:conf/acl/CaiCW19, DBLP:conf/acl/LiangLLY00PX22, DBLP:conf/aaai/QiaoJSCZN23}. \citeauthor{DBLP:conf/acl/CaiCW19} introduce a new dataset based on Twitter specifically tailored for multimodal sarcasm detection. In our paper, we refer to this dataset as \textbf{Sarcasm}.
            }
            \item{\textbf{Multimodal Relation Extraction (MRE)}:
            MRE involves identifying relations between two entities, supported by visual-text contexts \citep{DBLP:conf/mm/ZhengFFCL021, DBLP:conf/coling/0023HDWSSX22, DBLP:conf/aaai/Yuan0WL23}. To support research in this domain, \citeauthor{DBLP:conf/mm/ZhengFFCL021} introduce the Multimodal Neural Relation Extraction dataset (\textbf{MNRE}), which is manually labeled and serves as a valuable resource for MRE tasks.
            }
        \end{itemize}

        \section{Experimental Results and Analysis}
	\label{EXPERIMENTS}
        We evaluate the zero-shot performance of 45 models (as detailed in Table \ref{Table_different_models}) across 16 datasets (listed in Table \ref{Table_dataset_statistic}), utilizing 10 distinct instructions per model (as shown in Figures \ref{Figure_1_PuzzleVQA_different_instructions}, \ref{Figure_1_MSD_Multimodal_Instruction}, and \ref{Figure_2_text_instruction_for_different_tasks}). The related results in a total of 7200 (45 $\times$ 16 $\times$ 10) experimental results, each yielding an accuracy score denoted as $acc_{mdi}$ from Eq. \ref{eq:acc}.
        Although most of the evaluated datasets could rely on text context for inference, PuzzleVQA, and MMMU are exceptions, necessitating multimodal contexts. For ScienceQA, which primarily depends on the image modality to answer questions, we also explore the potential to extract knowledge from models using solely the image modality.
        Our comprehensive evaluation incorporates a range of metrics, including the Best Performance metric, the Mean Relative Gain metric, the Stability metric, and the Adaptability metric. Detailed analyses of these metrics and insightful conclusions related to them are presented in the subsequent sections.

        \subsection{Best Performance}
        \begin{table*}[t] 
        \small
        \begin{center}
        \renewcommand{\arraystretch}{1.3} 
        \caption{
        The best zero-shot performance, $A^{\tilde{i}}$ ($\uparrow$, measured by Accuracy), of various large language models on different datasets. Superscripts indicate the specific instruction that yields the best performance for each dataset on the respective model.
        The ``\textbf{Total}'' column represents the sum of accuracy scores across all datasets for each model. MVSA-S, MVSA-M, Twitter15, and Twitter17 refer to the MVSA-Single, MVSA-Multiple, Twitter-2015, and Twitter-2017 datasets, respectively. AdapterV2 and MultiGPT represent the LLaMA-AdapterV2 and Multimodal-GPT models, respectively.
        $\heartsuit$ indicates that the specific closed-source models (GPT-4V and Claude3-V) are evaluated on subsets.
        The highlighted \textbf{\textcolor[RGB]{34,139,34}{text}} indicates the best performance on \textbf{the full datasets}.
        The highlighted \textbf{\textcolor[RGB]{153,50,204}{text}} denotes superior performance on \textbf{the subsets}, outperforming the top results of the model evaluated across the full dataset.
        The notation ``-'' indicates datasets (PuzzleVQA and MMMU) where LLM assessment was not conducted, as these datasets inherently require visual and textual inputs for meaningful response generation. The ``\textbf{Total}'' denotes the aggregated performance across all evaluated datasets, while ``\textbf{\textit{Total$^\star$}}'' represents the aggregate results, excluding the VQAMC task (ScienceQA, PuzzleVQA, and MMMU).
        }
        % \vspace{-1em}
        \resizebox{1\textwidth}{!}{
        \begin{tabular}{
        p{3.7cm}< \centering| 
        p{1.3cm}< \centering p{1.3cm}< \centering p{1cm}< \centering |
        p{1.3cm}< \centering p{1.4cm}< \centering p{1.1cm}< \centering| 
        p{1.2cm}< \centering p{1.2cm}< \centering p{1.4cm}< \centering  p{1.4cm}< \centering| 
        p{1cm}< \centering p{1cm}< \centering p{1cm}< \centering|
        p{1cm}< \centering| p{1.1cm}< \centering| p{1cm}< \centering|
        p{1.1cm}< \centering| p{1cm}< \centering 
        }
        \toprule[1pt]
        \multirow{2}{*}{\textbf{Models}}
        & \multicolumn{3}{c|}{\textbf{VQAMC}} 
        & \multicolumn{7}{c|}{\textbf{MSA}} 
        & \multicolumn{3}{c|}{\textbf{MABSA}}
        & \textbf{MHMD} 
        & \textbf{MSD}
        & \textbf{MRE} & \multirow{2}{*}{\textbf{Total}} & \multirow{2}{*}{\textbf{\textit{Total$^\star$}}}\\
        \cline{2-17}
        & \textbf{ScienceQA} & \textbf{PuzzleVQA} & \textbf{MMMU}
        & \textbf{MVSA-S} & \textbf{MVSA-M} & \textbf{TumEmo} &  \textbf{MOSI-2} & \textbf{MOSI-7} & \textbf{MOSEI-2} & \textbf{MOSEI-7} 
        & \textbf{Twitter15} & \textbf{Twitter17} & \textbf{MASAD} 
        & \textbf{Hate} 
        & \textbf{Sarcasm} 
        & \textbf{MNRE}  & \multicolumn{1}{c|}{}\\
        \midrule[1pt]
        \textbf{\textcolor{red}{ChatGPT}} & 69.41$^7$ & - & - & 56.55$^3$  & 53.18$^7$ & 48.17$^3$ & 89.60$^5$ & 44.44$^{10}$ & 84.97$^5$ & 40.77$^1$ 
        & 65.48$^4$ & 59.97$^{10}$ & 72.70$^3$ & 60.84$^8$ & 69.02$^7$ & 38.28$^2$ &  - & \textit{783.97} \\
        \textbf{LLaMA1-7B} & 36.19$^5$ & - & - & 67.23$^1$ & 60.72$^1$ & 38.26$^1$ & 82.01$^2$ & 34.26$^2$ & 75.62$^1$ &  15.50$^1$
        & 58.53$^3$ & 46.43$^3$ & 65.67$^7$ & 50.40$^4$ & 58.99$^4$ & 2.66$^3$ &  -  & \textit{656.28} \\
        \textbf{LLaMA1-13B} & 43.33$^6$ & - & - & 66.99$^2$ & 68.82$^8$ & 44.68$^6$ & 72.10$^5$ & 34.11$^2$ & 79.55$^2$ & 28.74$^2$
        & 52.07$^3$ & 47.24$^7$ & 65.49$^2$ & 49.20$^5$ & 57.53$^4$ & 19.22$^5$ &  -  & \textit{685.74}\\
        \textbf{LLaMA2-7B} & 43.08$^6$ & - & - & 66.99$^1$ & 69.22$^2$ & 40.28$^4$ & 67.68$^1$ & 26.38$^1$ & 77.30$^1$ & 16.78$^1$ 
        & 58.53$^3$ & 46.60$^2$ & 67.19$^2$ &	52.00$^1$ & 56.33$^1$ & 3.59$^7$ &  - & \textit{648.87} \\
        \textbf{LLaMA2-13B} & 55.78$^7$ & - & - & 66.02$^1$ & 68.69$^7$ & 45.78$^6$ & 81.86$^2$ & 31.49$^6$ & 81.66$^2$ & 24.33$^6$ 
        & 60.37$^5$ & 48.54$^5$ & 69.10$^2$ & 55.00$^2$ & 60.23$^1$ & 20.00$^5$ &   - & \textit{713.07} \\
        \textbf{LLaMA3.1-8B-Instruct} & 60.34$^3$ & - & - & 62.14$^{10}$ & 61.94$^6$ & 46.08$^1$ & 87.31$^7$ & 34.94$^5$ & 75.12$^7$ & 31.30$^1$ & 50.63$^7$ & 50.81$^1$ & 70.15$^6$ & 58.40$^7$ & 56.04$^3$ & 38.12$^6$ & - & \textit{722.98}\\
        \textbf{Mixtral-AWQ} & 66.63$^3$ & - & - & 54.37$^9$ & 55.59$^{10}$ & 43.94$^1$ & 87.92$^5$ & 41.23$^7$ & 79.30$^5$ &  38.27$^7$ 
        & 55.45$^1$ & 60.21$^2$ & 70.56$^1$ & 58.00$^5$ &  64.38$^7$ & 37.03$^9$ & - & \textit{746.25}\\
        \textbf{Gemma} & 52.95$^5$ & - & - & 67.72$^8$ & 61.61$^9$ & 43.10$^2$ &  81.65$^7$ & 36.11$^7$ & 77.05$^2$ & 25.06$^7$ & 54.29$^5$ & 52.43$^5$ & 70.94$^5$ & 57.80$^1$ & 60.07$^2$ & 30.31$^2$ &  - & \textit{718.14} \\
        \textbf{Flan-T5-XXL} & 67.43$^9$ & - & - & 64.81$^4$ & 66.01$^4$ & 49.56$^3$ &  89.60$^{10}$  & 42.86$^6$ & 86.52$^6$ & 46.29$^6$ 
        & \textbf{\textcolor[RGB]{34,139,34}{72.13$^5$}} & 63.70$^3$ & 74.39$^8$ & 57.40$^7$ & 71.40$^9$ & 31.41$^9$ &  - & \textit{816.08}\\
        \midrule[1pt]
        % \textbf{\textcolor[RGB]{153,50,204}{} 紫色
        % \textbf{\textcolor[RGB]{34,139,34}{} 绿色
        \textbf{\textcolor{red}{GPT-4V$^{\heartsuit}$}} & 79.80$^{10}$ & 31.11$^3$ &  \textbf{\textcolor[RGB]{153,50,204}{57.78$^9$}} & 76.19$^5$ &  71.05$^4$ &  50.58$^3$ &   \textbf{\textcolor[RGB]{153,50,204}{90.91$^3$}} &  61.19$^3$ & 87.10$^8$ &   \textbf{\textcolor[RGB]{153,50,204}{49.44$^1$}} & 53.85$^8$ & 60.16$^2$ & 78.95$^3$ &   78.00$^8$ &  76.76$^9$ & 54.29$^8$ &  \textbf{\textcolor[RGB]{153,50,204}{1057.16}} &  \textit{\textbf{\textcolor[RGB]{153,50,204}{888.47}}}
        \\
        \textbf{\textcolor{red}{GPT-4o$^{\heartsuit}$}} & 87.68$^3$ &  \textbf{\textcolor[RGB]{153,50,204}{37.78$^9$}} & 52.87$^{10}$ & 71.43$^9$ &  \textbf{\textcolor[RGB]{153,50,204}{71.71$^5$}} &  \textbf{\textcolor[RGB]{153,50,204}{53.96$^5$}} & 89.39$^9$  &53.73$^3$ & 86.38$^8$ & 45.25$^4$ & 51.92$^4$ & 57.72$^4$ & 81.38$^2$ &  \textbf{\textcolor[RGB]{153,50,204}{80.00$^1$}}  &  70.12$^2$ & 54.29$^8$ & 1045.61 & \textit{867.28} \\
        \textbf{\textcolor{red}{Claude3-V$^{\heartsuit}$}} & 76.35$^5$ &  36.11$^{10}$ & 53.33$^7$ &   \textbf{\textcolor[RGB]{153,50,204}{80.95$^4$}} & 69.08$^5$ & 46.15$^{10}$ & 78.79$^1$ & 32.84$^1$ & 79.93$^7$ & 30.73$^1$ & 38.46$^5$ & 54.47$^5$ & 76.32$^1$ & 66.00$^5$ & 71.37$^2$ &   \textbf{\textcolor[RGB]{153,50,204}{57.14$^{10}$}} & 948.02$^{\heartsuit}$ & \textit{782.23}
        \\
        \textbf{\textcolor{red}{GLM-4V-plus$^{\heartsuit}$}} &  \textbf{\textcolor[RGB]{153,50,204}{95.07$^3$}} & 35.56$^6$ & 57.14$^{10}$ & 78.57$^9$ & 70.49$^6$ & 49.42$^5$ & 89.39$^5$  &51.47$^5$ &  \textbf{\textcolor[RGB]{153,50,204}{87.81$^7$}} & 48.88$^1$ &  \textbf{\textcolor[RGB]{153,50,204}{56.73$^7$}} &  \textbf{\textcolor[RGB]{153,50,204}{64.23$^7$}} &  \textbf{\textcolor[RGB]{153,50,204}{81.58$^8$}} & 66.00$^8$ &  \textbf{\textcolor[RGB]{153,50,204}{78.42$^7$}} & 38.57$^9$ & 1049.33 & \textit{861.56} \\
        \midrule[1pt]
        \textbf{\textcolor{red}{Gemini-V}} & 80.35$^5$ & 26.72$^7$ &  47.11$^7$ & 72.73$^3$ & 70.18$^3$ & 51.65$^3$ & 88.34$^8$ &  48.68$^8$ &  \textbf{\textcolor[RGB]{34,139,34}{87.14$^7$}} & 38.86$^7$ 
        & 54.51$^8$ & 59.32$^8$ &  81.02$^3$ &  \textbf{\textcolor[RGB]{34,139,34}{68.32$^4$}} & 56.83$^{10}$ & 41.41$^4$ & 973.17 & \textit{818.99} \\
        \textbf{OpenFlamingo} &  39.27$^5$ &  23.67$^5$  &  24.44$^7$ & 55.58$^7$ & 61.15$^7$ & 29.47$^9$ & 79.97$^7$ & 24.85$^2$ & 77.30$^7$ & 12.12$^2$ 
        & 57.28$^5$ & 46.19$^5$ & 66.91$^7$ &	49.40$^2$ & 52.68$^1$ & 3.12$^6$ & 703.40 & \textit{616.02}\\
        \textbf{Fromage} & 34.51$^7$ & 21.94$^7$ & 20.33$^7$ & 29.85$^6$ & 28.19$^2$ & 22.76$^1$ & 57.19$^7$ & 19.15$^2$ & 47.41$^2$ & 11.04$^2$
        & 19.96$^6$ & 27.31$^6$ & 35.10$^6$ &	37.60$^2$ & 40.68$^7$ & 0.16$^1$ & 453.18 &   \textit{376.40} \\
        \textbf{LLaVA-v0-7B} & 41.10$^5$ &  19.94$^5$ & 24.85$^4$ & 69.42$^3$ & 65.42$^6$ & 30.44$^8$ & 74.69$^2$ & 30.03$^9$ & 74.65$^7$ & 18.12$^9$
        & 35.10$^3$ &  44.57$^3$ & 73.48$^3$ & 39.80$^5$ &  43.21$^{10}$ &  4.06$^8$ &  688.88 &  \textit{602.99} \\
        \textbf{LLaVA-v0-13B} & 47.74$^7$ & 12.61$^2$ &  15.18$^7$ & 73.06$^7$ & 69.61$^2$ & 38.51$^4$ &  80.18$^7$ & 30.90$^6$ &  76.58$^7$ & 28.37$^3$
        & 37.99$^5$ & 48.46$^5$ & 77.69$^9$ & 40.40$^5$ &  44.29$^5$ & 3.91$^5$ &  725.48 &  \textit{649.95}  \\
        \textbf{LLaVA-v1.6-7B} & 65.89$^5$ & 28.11$^4$ & 34.76$^7$ & 59.95$^2$ & 67.23$^2$ & 45.12$^1$ & 85.63$^9$ & 30.41$^6 $ & 81.62$^1$ & 23.97$^7$ & 59.31$^9$ & 52.84$^1$ & 75.18$^9$ & 57.80$^2$ & 59.61$^5$ & 17.19$^1$ &  844.62 &  \textit{715.86}\\
        \textbf{LLaVA-v1.6-13B} & 69.66$^5$ & 26.89$^9$ & 39.53$^3$ & 64.56$^3$ & 60.43$^3$ & 53.22$^3$ & 86.39$^3$ & 35.96$^9$ & 78.26$^3$ & 40.89$^7$ & 58.73$^7$ & 56.08$^{10}$ & 77.95$^{10}$ & 61.20$^4$ & 62.31$^3$ & 22.81$^3$ &  894.87 &  \textit{758.79}\\
        \textbf{MiniGPT4} & 58.70$^5$ & 27.33$^5$ & 30.56$^5$ & 71.12$^5$ & \textbf{\textcolor[RGB]{34,139,34}{70.78$^2$}} & 50.29$^4$ & 83.99$^4$ & 35.42$^2$ & 83.38$^2$ &  38.46$^5$  
        & 47.16$^7$ & 49.43$^9$ & 76.01$^2$ & 54.20$^{10}$ & 57.49$^3$ &  2.81$^6$ & 837.13 &  \textit{720.54}\\
        \textbf{mPLUG-Owl} &  37.93$^7$ & 26.17$^3$ & 26.22$^6$ &  51.94$^6$ & 50.36$^6$ & 33.37$^1$ & 68.75$^1$ & 28.28$^6$ & 58.10$^6$ &  20.29$^6$
        & 33.75$^2$ & 38.74$^2$ & 58.26$^2$ & 48.60$^1$ &  49.73$^7$ &  5.47$^5$ & 635.96 &  \textit{545.64}\\
        \textbf{mPLUG-Owl2.1} & 60.29$^5$ & 25.61$^1$ & 33.33$^2$ & 53.64$^7$ & 63.11$^7$ & 47.02$^5$ & 85.63$^5$ & 35.67$^5$ & 73.47$^7$ & 33.92$^6$ & 60.66$^5$ & 55.11$^5$ & 75.60$^5$ &  58.20$^5$ & 60.48$^5$ & 22.19$^3$ & 843.93 &  \textit{724.70} \\
        \textbf{AdapterV2} &  54.44$^7$ & 25.72$^7$ & 27.11$^3$ & \textbf{\textcolor[RGB]{34,139,34}{73.54$^7$}} & 70.13$^4$ & 39.14$^{10}$  & 86.43$^8$ & 38.34$^8$ & 82.02$^8$ & 33.53$^9$ 
        & 37.32$^9$ & 48.38$^9$ & 76.29$^4$ & 50.40$^7$ &  57.20$^5$ &   7.19$^3$ &  807.18 & \textit{699.91} \\
        \textbf{VPGTrans} &  47.00$^3$ & 19.11$^3$ & 26.33$^3$ & 64.32$^3$ & 69.54$^3$ & 46.17$^9$ & 76.22$^4$ & 30.47$^4$ & 76.76$^4$ & 38.27$^6$ 
        & 42.62$^6$ & 44.81$^6$ & 74.14$^9$ & 46.00$^3$ &   56.04$^4$ &  2.50$^9$ & 760.30 &  \textit{667.86}\\
        % \textbf{Otter} & 30.58 & 7.78 & 7.78 & 10.90 & 13.07 & 35.81 &	49.2 & 39.81 & 2.81 \\
        \textbf{MultiGPT} & 36.29$^5$ & 23.93$^3$ & 21.00$^7$ & 52.91$^7$ & 62.03$^1$ & 30.26$^2$ & 68.35$^7$ & 25.58$^2$ & 72.76$^7$ & 10.17$^5$ 
        & 58.53$^5$ & 46.35$^5$ & 67.58$^7$ &	49.80$^2$ & 59.82$^4$ & 2.81$^1$ &  687.63 &  \textit{606.95} \\
        \textbf{LaVIN-7B} & 75.11$^3$ & 25.67$^3$ & 22.89$^2$ & 39.32$^2$ & 40.75$^2$ & 26.84$^7$  & 71.41$^5$ & 25.73$^5$ & 69.97$^7$ & 29.46$^1$
        & 37.22$^1$ & 33.06$^1$ & 60.08$^7$ & 50.40$^7$ & 60.48$^7$ & 12.34$^5$ & 680.73 & \textit{557.06}\\
        \textbf{LaVIN-13B} & 77.54$^5$ & 27.67$^2$ & 25.22$^7$ & 53.64$^4$ & 48.79$^4$ & 32.77$^4$ & 79.97$^7$ & 27.63$^1$ & 73.54$^7$ & 27.20$^7$ 
        & 35.39$^6$ & 40.68$^6$ & 62.76$^6$ &	49.60$^1$ & 57.58$^7$ & 11.56$^1$ & 731.54 &  \textit{601.11}\\
        \textbf{Lynx} & 38.28$^7$ & 22.72$^5$ & 19.22$^5$ & 64.32$^7$ & 67.71$^9$ & 42.79$^6$ & 74.77$^7$ & 22.37$^2$ & 73.72$^7$ & 10.28$^2$ & 46.00$^6$ & 47.00$^{10}$ & 73.52$^2$ &	51.60$^7$  & 43.96$^7$ &  9.22$^9$ & 707.48 & \textit{627.26} \\
        \textbf{Fuyu-8B} & 45.71$^5$ & 27.06$^{10}$ & 26.14$^5$ & 48.54$^7$ & 55.46$^7$ & 46.34$^9$  & 83.49$^3$ & 17.10$^3$ &  78.37$^3$ & 40.52$^9$
        & 58.82$^2$ & 50.81$^5$ & 74.53$^3$ & 53.00$^1$ &   61.44$^2$ & 1.72$^7$ &  769.05 &  \textit{670.14}\\ %decoder
        \textbf{LaVIT} & 42.09$^5$ & 27.22$^3$ & 26.56$^3$ & 61.65$^2$ & 68.74$^2$ & 41.78$^2$ & 73.09$^8$ & 28.95$^2$ & 64.10$^2$ &  15.50$^2$ 
        & 36.84$^1$ & 43.36$^2$ & 65.07$^4$ & 51.40$^6$ &  56.00$^3$ &    0.62$^3$ & 702.97 &  \textit{607.10}\\ %decoder
        \textbf{LLaMA-3.2-11B-Vision} & 86.12$^5$ & 26.39$^2$ & 46.11$^5$ &  63.11$^5$ & 63.31$^6$ & 52.08$^5$ & 85.32$^5$ & 38.16$^1$ & 78.80$^7$ &  34.39$^{10}$ & 59.50$^7$ &  56.97$^1$ & 77.18$^{10}$ & 58.60$^2$ & 57.04$^5$ & 34.69$^{10}$ &  917.77 &   \textit{759.15}\\
        \textbf{MiniCPM-V2.6} & 93.11$^5$ & 28.17$^4$ &  42.27$^3$ &  70.39$^4$ &  70.18$^{10}$ &  52.80$^1$ &   89.14$^{10}$ &  44.74$^{10}$ &  83.63$^{6}$ &  43.90$^1$ &   44.65$^7$ &  52.84$^7$ &  79.74$^{10}$ &  64.40$^{10}$ &   67.16$^7$ &  42.66$^5$ &  969.78 &  \textit{806.23} \\
        \textbf{BLIP-2} & 74.17$^1$  & 28.00$^5$ & 35.22$^{10}$ & 66.26$^4$ & 68.22$^9$ & 51.06$^3$ & 88.99$^9$ & 43.42$^2$ & 86.88$^6$ & 45.79$^6$
        & 70.78$^5$ & \textbf{\textcolor[RGB]{34,139,34}{64.42$^3$}} & 77.59$^9$ &	58.00$^7$ & 72.02$^2$ & 34.69$^1$ & 965.51 & \textit{828.12} \\
        \textbf{InstructBLIP} & 73.33$^2$  & 28.67$^7$ & 33.44$^6$ & 71.60$^6$ & 70.37$^6$ & 52.36$^8$ & 88.68$^9$ & 43.28$^2$ & 85.98$^9$ & 45.68$^9$ 
        & 63.07$^5$ & 62.72$^3$ & 80.53$^{10}$ &	58.20$^9$ & 73.10$^7$ & 36.72$^2$ &  967.73 & \textit{\textbf{\textcolor[RGB]{34,139,34}{832.29}}} \\
        \textbf{GLM-4V-9B} & \textbf{\textcolor[RGB]{34,139,34}{97.03$^5$}} & 29.33$^3$ & 43.67$^5$ & 70.87$^5$ & 70.31$^{10}$ & 52.56$^3$ & 87.61$^2$ & 40.35$^3$ & 81.37$^7$ & 45.43$^2$ & 52.36$^4$ & 53.65$^4$ & 81.34$^6$ & 63.20$^4$ &  60.52$^7$ & 34.53$^5$ &  964.13 &  \textit{794.10} \\ 
        \textbf{InternVL2.5-8B} & 95.14$^5$ & 27.06$^5$ & \textbf{\textcolor[RGB]{34,139,34}{47.44$^5$}} & 67.48$^6$ & 67.04$^6$ & 49.68$^3$ & 86.09$^5$ & 41.52$^7$ & 80.48$^5$ & 42.36$^2$ & 61.04$^9$ & 59.16$^9$ & 79.31$^1$ & 63.60$^8$ &  67.04$^5$ & 47.44$^5$ & 981.88 &   \textit{812.24}\\
        \textbf{DeepSeek-VL2-tiny} & 76.65$^5$ &  25.83$^5$ &  35.33$^5$ &  72.82$^7$ &  70.70$^7$ &   45.40$^1$ &   83.64$^3$ &  33.92$^5$ &  83.20$^3$ &   27.62$^9$ &  55.16$^6$ &  53.40$^6$ &  78.34$^6$ &  53.40$^4$ &   65.42$^2$ &  17.03$^7$ &  877.86 &  \textit{740.05} \\
        \textbf{DeepSeek-VL2-small} & 71.66$^5$ & 26.32$^4$ & 35.22$^3$ & 61.17$^2$ & 60.76$^2$ & 40.54$^{10}$ & 80.28$^4$ & 24.85$^2$  &74.04$^4$ & 39.97$^4$ & 57.86$^7$ & 48.59$^7$ &  76.25$^2$ & 57.60$^4$ &  60.61$^{10}$ &  8.91$^5$ &  824.63 &   \textit{691.43}\\
        \textbf{DeepSeek-VL2} & 70.80$^5$ &  26.44$^6$ & 40.22$^5$ & 68.45$^{10}$ & 67.69$^{10}$ & 48.01$^5$ & 86.09$^5$ & 36.84$^3$ & 84.02$^2$ & 40.75$^6$ & 65.96$^5$ & 56.89$^3$ & 72.00$^{10}$ & 57.20$^3$ &  62.31$^7$ & 29.53$^3$ &  913.2 &  \textit{775.74}  \\ 
        \textbf{Qwen-VL-Chat} & 63.51$^3$ & 27.83$^6$ & 35.33$^5$ & 62.38$^6$ & 69.06$^6$ & 49.29$^1$ & 85.93$^7$ & 33.92$^6$ & 80.41$^7$ & 44.12$^1$
        & 65.48$^7$ & 59.72$^4$ & 73.74$^7$ & 54.40$^1$ &  61.10$^7$ &  20.47$^3$ &  886.69 &  \textit{760.02}   \\
        \textbf{Qwen2-VL-7B-Instruct} & 82.76$^5$ & 30.06$^4$ &  46.11$^5$ &  55.83$^1$ &  61.74$^1$ &  50.92$^7$ &  87.46$^6$ &  41.64$^5$ &  84.27$^1$ &  45.29$^6$ &  65.19$^5$ &  60.37$^5$ &  78.89$^5$ &  66.40$^8$ &   66.21$^7$ &  35.62$^4$  &   958.76 &  \textit{799.83}\\
         \textbf{Qwen2-VL-72B-Instruct} & 71.94$^5$ & 29.94$^5$ & 44.89$^5$ & 55.10$^6$ &  58.14$^6$ & 50.26$^{10}$ & \textbf{\textcolor[RGB]{34,139,34}{90.52$^9$}} & \textbf{\textcolor[RGB]{34,139,34}{49.56$^5$}} & 83.55$^9$ & \textbf{\textcolor[RGB]{34,139,34}{47.63$^4$}} & 68.18$^5$ & 62.48$^1$ & 74.43$^6$ & 61.80$^{10}$ &  
        \textbf{\textcolor[RGB]{34,139,34}{76.17$^2$}} & \textbf{\textcolor[RGB]{34,139,34}{52.19$^6$}}  & 976.78 &  \textit{830.01} \\
        \textbf{Qwen2.5-VL-3B-Instruct} & 78.78$^5$ & 30.83$^3$ & 42.89$^7$ & 69.90$^3$ &  68.35$^3$ & 51.98$^5$ & 87.92$^3$ & 40.94$^6$ & 78.55$^3$ & 46.32$^6$ & 64.32$^7$ & 59.97$^7$ & 80.18$^9$ & 60.00$^1$ &   62.76$^6$ & 28.28$^3$ &  951.97 & \textit{799.47}\\
        \textbf{Qwen2.5-VL-7B-Instruct} & 87.36$^5$ & \textbf{\textcolor[RGB]{34,139,34}{31.17$^5$}} & 44.91$^3$ & 68.45$^5$ & 66.91$^5$ & \textbf{\textcolor[RGB]{34,139,34}{53.77$^3$}} & 89.14$^7$ & 42.11$^7$ & 85.56$^5$  &45.96$^8$ & 64.42$^7$ & 61.26$^6$ & \textbf{\textcolor[RGB]{34,139,34}{81.38$^9$}} & 64.00$^{10}$ &   64.30$^6$ &  44.22$^5$ &  \textbf{\textcolor[RGB]{34,139,34}{994.92}} &  \textit{831.48} \\
        \bottomrule[1pt]
        \end{tabular}
        }
        \label{Best_Acc}
        \end{center}
        % \vspace{-1em}
        \end{table*}

        We present the best performance ($A^{\tilde{i}}$), calculated using Eq. \ref{equ_best_acc}, achieved by different models across various multimodal datasets. This metric serves as a set of benchmarks for evaluating the performance of different models on each dataset. Based on these results, we make the following observations.
        % \vspace{-1em}
        \subsubsection{Performance of Various Models on Full Datasets}
        We first report the best performance of various models across full datasets, as detailed in Table \ref{Best_Acc}. Many later models were trained on the ScienceQA dataset, while the MMMU and PuzzleVQA datasets require models to utilize visual context to answer questions. For this reason, we also calculate the total result for each model excluding the VQAMC task (\textit{Total$^{\star}$}).
        
\begin{figure*}[t] %%
          \centering %?????????
          \includegraphics[width = 1\textwidth]{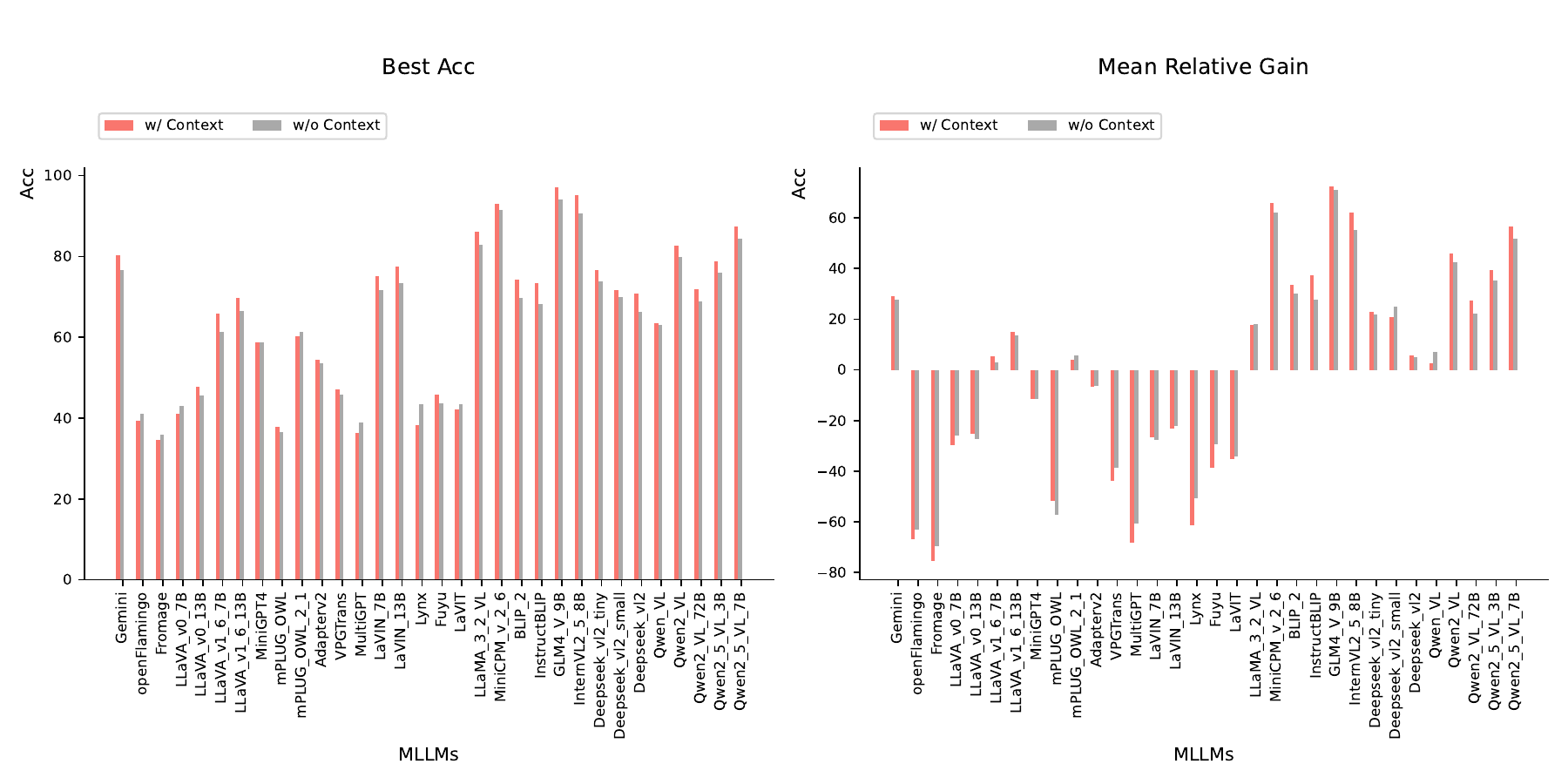}
          %\vpsace{-1em}
          \caption{Comparison of experimental results in ScienceQA (with/without Text Context) for best performance, (measured by Accuracy), and Mean Relative Gain of MLLMs across all instructions. `w/' means with and `w/o' means without.
           }
          \label{ScienceQA_with_or_without_context} 
          %\vpsace{-1.5em}
        \end{figure*}

        (1) \textbf{Comparison between open-source models and closed-source models}:
        Closed-source models such as ChatGPT, GPT-4V, and GPT-4o show excellent performance on various evaluation metrics. Among the open-source alternatives, the Flan-T5 series (including Flan-T5, BLIP-2, and Instruct-BLIP) and Qwen2.5-VL models (such as Qwen2.5-VL-3B-Instruct and Qwen2.5-VL-7B-Instruct) show excellent capabilities in multimodal reasoning tasks, but there is still a gap compared to closed-source models with excellent performance. There are two possible reasons for this performance gap. The first reason is training data uncertainty. We cannot be sure whether the closed-source models are trained on our evaluation dataset, which can explain their excellent performance. Another reason is the model size difference. The architecture and parameters of closed-source models may be much larger compared to open-source alternatives. However, we cannot verify this due to limited access to their architecture specifications and training details.
        (2) \textbf{Flan-T5 backbone vs. LLaMA backbone in the open-source models}:
        Among open-source models, except for the VQAMA task, InstructBLIP stands out by achieving top performance on three datasets and securing the first position in the `\textit{Total$^{\star}$}' column in the Table \ref{Best_Acc}. Notably,  InstructBLIP leverages Flan-T5-XXL as its backbone model. 
        The superior performance of Flan-T5-XXL can be attributed to its Encoder-Decoder architecture, which contrasts with the Decoder-only architecture of the LLaMA series models. The encoder module in Flan-T5-XXL offers significant advantages in multimodal representation learning, making it particularly well-suited for addressing multimodal reasoning tasks with complex multimodal contexts.
        (3) \textbf{LLMs vs. MLLMs}:
        For the cumulative best performance of models across all datasets—excluding the VQAMA task, as indicated by the ``\textbf{\textit{Total$^\star$}}'' column—we observe that LLMs exhibit superior performance compared to certain MLLMs, such as Openflamingo, Fromage, LaVIT, and  others.
        The phenomenon may be due to theses MLLMs tend to specialize in traditional visual language (VL) capabilities, including language generation, recognition, knowledge reasoning, spatial reasoning, and OCR, linked to the image modality. Meanwhile, these models might not sufficiently prioritize capabilities like multimodal alignment, multimodal fusion, and etc. Such capabilities are critical for effectively handling multimodal reasoning tasks that involve both vision and text contexts. 
        (4) \textbf{Different scale checkpoints}:
        In the same model series, larger versions, evidenced by increased parameters such as in LLaMA-7B vs. LLaMA-13B, LLaVA-v0-7B vs. LLaVA-v0-13B, and LaVIN-7B vs. LaVIN-13B, generally outperform their smaller counterparts across most datasets. It suggests that models with more parameters and extensive training are likely to achieve superior performance. 
        (5) \textbf{Latest models}:
        The updated model iterations, such as the LLaMA, Qwen-VL, mPLUG-OWL, and LLaVA series, demonstrate significant improvements over their predecessors, highlighting how advancements in novel architectures and extended training contribute substantially to enhanced model efficacy. Additionally, the latest models, such as Qwen2.5-VL and MiciCPM-V2.6, showcase exceptional performance. Remarkably, newer models with fewer parameters (e.g., Qwen2.5-VL-7B-Instruct) can match the performance of earlier models with significantly more parameters (e.g., Qwen2-VL-72B-Instruct), underscoring the higher information density of the latest models.
        (6) \textbf{Various instructions}:
        On one hand, performance peaks vary within the same model when applied to different datasets, contingent on the instructions used. On the other hand, across various models, the optimal performance on a specific dataset often requires model-specific instructions. Despite its importance, the area of selecting the most suitable instructions for diverse datasets and models is under-researched and represents a promising avenue for future exploration.
        (7) \textbf{More challenge tasks}:
        Closed-source models, such as Gemini-V, significantly outperform open-source models on more challenging datasets, including MOSI-7 and MOSEI-7, which involve multiple classification categories, MNRE with a long-tail effect, and complex reasoning datasets like PuzzleVQA and MMMU. These results highlight considerable room for improvement in open-source models when addressing such tasks.
        Closed-source models demonstrate superior performance on sophisticated tasks like PuzzleVQA and multimodal relation extraction for several technical reasons. 1) Advanced architectural design: These models likely employ more sophisticated attention mechanisms and deeper neural architectures that better handle complex reasoning chains and long-range dependencies. 2) Higher-quality training data: Their training likely involves carefully curated datasets specifically targeting complex reasoning patterns and edge cases. 3) More effective pre-training strategies: They possibly utilize advanced techniques like chain-of-thought prompting and multi-task learning during pre-training to develop stronger reasoning capabilities. 4) Broader knowledge integration: The larger scale of training data and parameters enables better integration of world knowledge, which is crucial for handling long-tail phenomena and multi-step reasoning. 5) Better optimization techniques: They may benefit from more sophisticated training objectives and optimization strategies that specifically enhance performance on complex reasoning tasks.
        Nevertheless, the latest open-source models have shown remarkable progress in addressing these complex challenges. Notably, the Qwen2.5-VL-7B-Instruct model exhibits competitive performance on sophisticated tasks like PuzzleVQA and MNRE. This progress suggests that open-source models are gradually narrowing the performance gap with closed-source counterparts through improved architectural designs, more effective training strategies, and enhanced multimodal integration capabilities.
        (8) \textbf{The inference speed}: 
        Notably, models using LLaMA as their backbone exhibit slower inference speeds compared to those based on FLAN-T5 and Qwen architectures. This performance difference stems primarily from LLaMA's higher parameter density and the lack of specialized inference optimizations that are present in FLAN-T5 and Qwen-based models.
        (9) \textbf{Data leakage}:
        Recent multimodal language models have demonstrated remarkable performance on ScienceQA, with MiniCPM, InternVL2.5, and GLM-4V-9B standing out as particularly noteworthy. This superior performance can be attributed to the explicit training of these models on the ScienceQA dataset. In contrast, other models, such as the Qwen-VL series, LLaMA-3.2-Vision-Instruct, and DeepSeek-VL series, lack evidence of similar training, as indicated in their related papers.
        This observation raises broader concerns about potential data leakage issues, which may not be limited to the ScienceQA dataset.
        (10) \textbf{With/Without text context on ScienceQA}:
        ScienceQA primarily relies on the vision context to answer questions. To investigate this further, we conduct comparative experiments excluding textual context, as detailed in the ``Text context'' section of Figure \ref{Figure_1_PuzzleVQA_different_instructions}. The results of these experiments are presented in Figure \ref{ScienceQA_with_or_without_context}.
        Our observations indicate that while most MLLMs exhibit improved performance on the ScienceQA dataset when textual context is included, the degree of improvement is relatively limited. This finding suggests that the presence or absence of textual context has only a minimal impact on the overall performance of the ScienceQA dataset.

        \subsubsection{Full datasets vs. Subsets}
\begin{figure*} %%
          \centering %?????????
          \includegraphics[width = 0.94\textwidth]{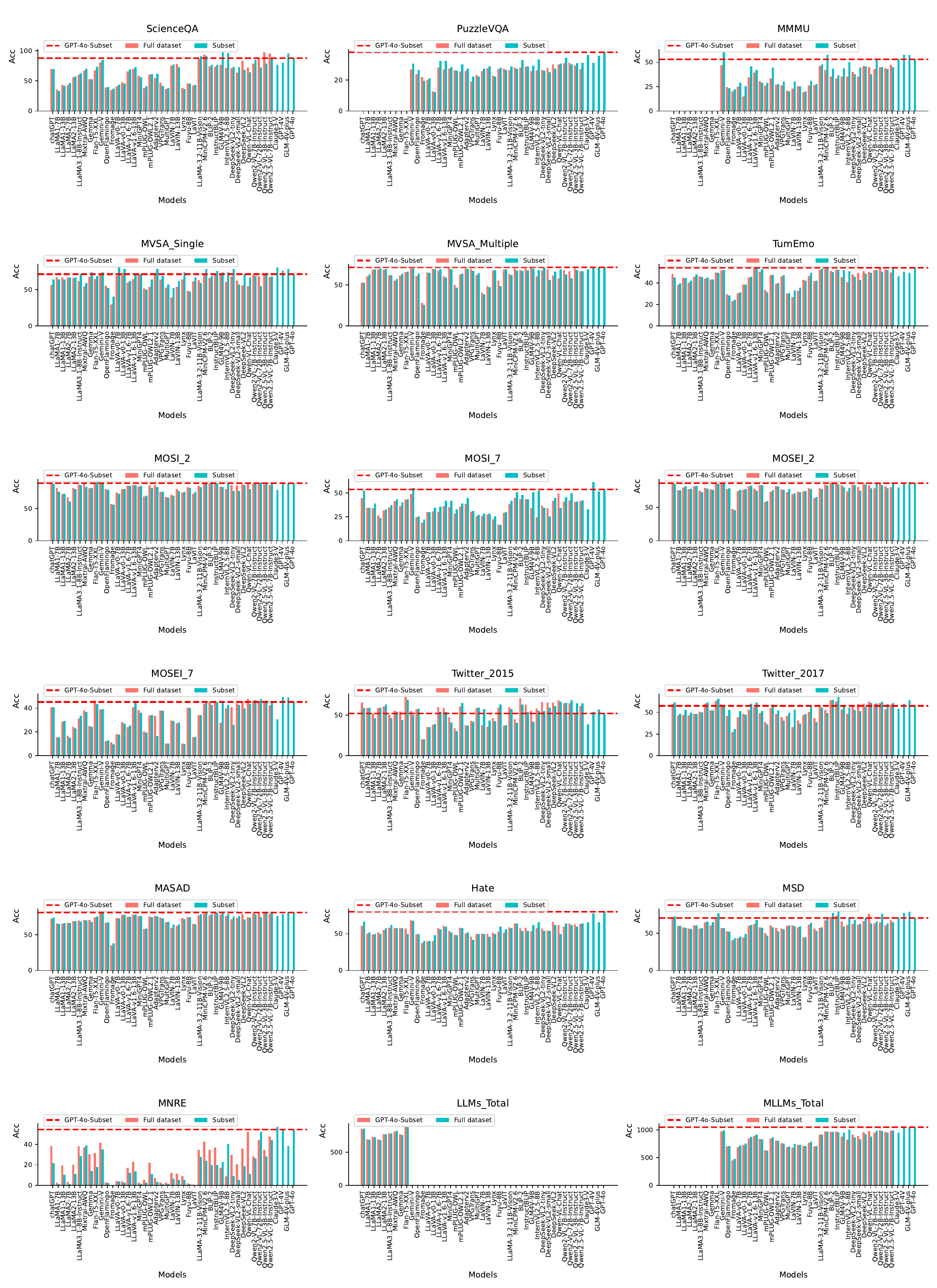}
          %\vpsace{-1em}
          \caption{Comparing results of the best performance metric between the full test dataset (\textbf{\textcolor[RGB]{249,118,110}{Full dataset}}) and a subset of the test dataset (\textbf{\textcolor[RGB]{0,191,196}{Subset}}). GPT-4V only has results on a subset, as indicated by the red dashed line (\textbf{\textcolor{red}{GPT-4V-Subset}}). The ``\textbf{Total}'' represents the sum of accuracy scores across all datasets for each model.
           }
          \label{Figure_GPT-4V_vs_other_models} 
          %\vpsace{-1.5em}
        \end{figure*}

        As discussed in \ref{subsubsection:MLLLMs}, given the high cost and limited availability of certain closed models, we draw inspiration from Consistently Distributed Sampling (CDS) \citep{yang2023few_MultiPoint} to tailor our evaluation. Following this, we exclusively evaluate GPT-4V, GPT-4o, Claude3-V, and GLM-4V-plus on these subsets across all tasks, selecting corresponding experimental results from other models for comparison. The comparative analysis, highlighting the best performance metrics between the full dataset and its subset, is illustrated in Figure \ref{Figure_GPT-4V_vs_other_models}.
        The MMMU and PuzzleVQA datasets require models to utilize visual context to answer questions, so for these two datasets we only test the performance on MLLMs. Therefore, we do not test MMMU and PuzzleVQA datasets on 9 LLMs. Therefore, we use the ``LLMs-Total'' metric to calculate the sum of the best accuracies across 14 datasets (excluding PuzzleVQA and MMMU) to measure each LLM's overall performance. Similarly, the ``MLLMs-Total'' metric sums up the best performances across all 16 datasets, including PuzzleVQA and MMMU to evaluate MLLM performance. 
        
        The subplots ``LLMs-Total'' and ``MLLMs-Total'' reveal minimal performance differences between models on full datasets and their subsets. This indicates that our evaluation of models like GPT-4V, GPT-4o, and other closed-models on subsets provides an accurate reflection of their overall performance across tasks.
        Furthermore, GPT-4o consistently achieves superior performance on most tasks. It excels particularly in challenging datasets with extensive classification categories, such as MOSI-7 and MOSEI-7, datasets with long-tail effects like MNRE, and complex reasoning tasks such as PuzzleVQA and MMMU, where it significantly outperforms other models. Notably, some open-source models, including InstructBLIP, BLIP-2, Qwen2.5-VL, and MiniCPM-V-2.6, also deliver competitive results across many subsets.
        An additional observation is that larger datasets exhibit less variation in experimental outcomes. This stability underscores the value of using subsets from large datasets, as they tend to provide more representative and reliable results.

        \begin{table*}[t]  \small
        \begin{center}
        \renewcommand{\arraystretch}{1.3}
        \caption{The mean relative gain, MRG$^{\mathcal{M}}$ ($\uparrow$), for various models across instructions.
        Positive values indicate performance above the average, while negative values indicate performance below the average. The `Wins1' and `Wins3' columns display the number of tasks where a model achieves the highest aggregated performance and the top 3 aggregated performance, respectively.
        Text in `\textbf{\textcolor[RGB]{34,139,34}{bold}}', `\underline{\textbf{underline}}', and `\textbf{$\star$}'  indicate the best scores, the sub-optimal scores, and the third-best scores, respectively.
        The notation ``-'' signifies that we do not assess LLMs for PuzzleVQA and MMMU datasets. The ``Win1'' and ``Win3'' results of LLMs are also not included.
        }
        % \vspace{-1em}
        \resizebox{1\textwidth}{!}{
        \begin{tabular}{
        p{3.6cm}< \centering| 
        p{1.3cm}< \centering p{1.3cm}< \centering p{1.1cm}< \centering |
        p{1.3cm}< \centering p{1.4cm}< \centering p{1cm}< \centering| 
        p{1.2cm}< \centering p{1.2cm}< \centering p{1.4cm}< \centering  p{1.4cm}< \centering| 
        p{1cm}< \centering p{1cm}< \centering p{1cm}< \centering|
        p{1cm}< \centering| p{1.1cm}< \centering| p{1cm}< \centering|
        p{0.5cm}< \centering p{0.6cm}< \centering
        }
        \toprule[1pt]
        \textbf{Models} 
        & \textbf{ScienceQA} & \textbf{PuzzleVQA} & \textbf{MMMU}
        & \textbf{MVSA-S} & \textbf{MVSA-M} & \textbf{TumEmo} &  \textbf{MOSI-2} & \textbf{MOSI-7} & \textbf{MOSEI-2} & \textbf{MOSEI-7} 
        & \textbf{Twitter15} & \textbf{Twitter17} & \textbf{MASAD} & 
        \textbf{Hate} & \textbf{Sarcasm} & \textbf{MNRE} & 
        \textbf{Wins1} & \textbf{Wins3}\\
        \midrule[1pt]
        \textbf{\textcolor{red}{ChatGPT}} &  61.41 & - & - &  2.21 &   1.33 &  33.34 &  37.53 &  58.55 &  34.77 &  74.52 &  82.43 &  52.06 &  45.26 &  28.13 &  30.68 &  79.58 & - & -\\
        % & 60.65  & 0.33 & -2.47 & 40.44 
        % & 48.48$^{\star}$ & \textbf{\textcolor[RGB]{34,139,34}{83.14}} & 42.00   & 82.38 
        % & \textbf{\textcolor[RGB]{34,139,34}{88.90}} &  57.01 & 35.43 & 38.33 & 39.09 & 215.60 & \textbf{2} & \textbf{3}  \\
        \textbf{LLaMA1-7B} &  -48.40 & - & - &  -48.88 & -48.63 & -37.10 &  -25.66 & -28.66  &-28.45 & -53.58 & -44.28 & -46.89 &  -46.30 &  -17.31 & -26.08 & -95.75 & - & -\\
        \textbf{LLaMA1-13B} & -51.18 & - & - & -18.16 & -12.29 & -39.41 & -24.07 & -25.46  &-14.96 & -42.92 & -25.19 & -18.97 &  -16.05 & -23.89 & -35.26 & -75.26 & - & - \\
        \textbf{LLaMA2-7B} &  -35.38 & - & - &  -6.07 &  -2.89 & -26.21 & -33.90 &  -46.06  &-11.52 & -57.28 & -43.07 & -37.23 &  -36.11 & -33.83 & -22.16 & -92.47 & - & - \\
        \textbf{LLaMA2-13B} &-47.46 & - & - & -23.44 & -23.19 & -17.15 & -25.69 & -44.34 & -29.69 & -61.47 & -40.84 & -39.93 &  -44.08 & -35.05 & -32.39 & -82.78 & - & - \\
        \textbf{LLaMA3.1-8B-Instruct}&  -7.66 & - & -  &  8.39 &   7.63 &  -2.91 &  12.07  & -6.98  &  2.63 & -14.61 &   0.28  &  6.81  &  21.30 &   15.18 &   2.52  & 97.95 & - & - \\
        \textbf{Mixtral-AWQ}  & 34.42 & - & -   & 3.88  &  0.19  & 16.84 &   6.79  & 27.87   & 1.14  & 56.47 &   7.64 &   8.89 &   0  &   18.73 &  25.29 &  50.85 & - & - \\
        \textbf{Gemma} &  20.33 & - & - &  \textbf{\textcolor[RGB]{34,139,34}{47.36}}  & 38.92 &  25.60 &   14.90 &   13.58  &  2.21 & -13.75 &  12.68 &  22.13  &  23.09  & 17.54 &   7.43 &  35.41 & - & - \\
        \textbf{Flan-T5-XXL} & \underline{\textbf{73.93}} & - & -  & 34.72 &  38.92 &  \textbf{47.00}$^\star$ &    \textbf{38.02}$^\star$ &  51.50 & 43.86 & \textbf{112.63}$^\star$  & 50.35 &  \textbf{53.13}$^\star$ &   \textbf{\textcolor[RGB]{34,139,34}{52.89}} &  30.50  &  \textbf{49.99}$^\star$ &  82.47 & - & - \\
        \midrule[1pt]
        \textbf{\textcolor{red}{Gemini-V}} & 29.10 &  -20.14 &  14.17 &  40.74 &  10.52 &  29.98 &  24.20 &   45.47 &  \textbf{50.42}$^\star$  & -3.20 & -3.42 &  21.42 &  \underline{\textbf{34.58}} &  \textbf{\textcolor[RGB]{34,139,34}{66.87}} &  22.32 &  94.82 & \textbf{1} & \textbf{3} \\
        \textbf{OpenFlamingo}  & -66.71 & -53.38 & -57.18 & -51.64 & -58.45 & -51.11 & -38.71 & -63.69 & -36.50 & -78.28 &  -63.44 & -57.15 & -51.35 & -33.70 &  -36.56 & -95.11 & 0 & 0 \\
        \textbf{Fromage}  & -75.47 & -72.43 & -68.42 & -74.13 & -88.49 & -81.48 & -70.79 & -84.53 & -71.88 & -92.83 & -87.20 &  -82.27 & -75.26 & -52.95 & -45.58 & -99.66 & 0 & 0 \\
        \textbf{LLaVA-v0-7B}  &  -29.85 & -46.48 & -41.62 &  20.76 & -12.51 & -45.06 &  -6.98 & -13.52 &  10.12 & -52.97 & -45.30 &  -24.01 &   3.93 & -49.10 &  -17.35 & -89.40 & 0 & 0 \\
        \textbf{LLaVA-v0-13B} & -25.15 & -67.29 & -65.52 &  33.36 &   1.49 &  -7.22 &  -6.94 & -26.41 &  13.22 & -40.87 &  -31.14 &  -5.83  & 27.92 & -24.21 & -15.23 & -87.26 & 0 & 0\\
        \textbf{LLaVA-v1.6-7B} &  5.38 & -17.94 & -12.19 &  12.46 &   0.29 &  20.65 &  19.31 &  -8.63 &  41.41 & -47.10 & -3.13 &  10.89 &  28.34 & 52.32 &  45.68 & -54.26 & 0 & 0\\
        \textbf{LLaVA-v1.6-13B} & 14.82 & -20.13 &   2.57 &  19.99 & -8.33 &  \textbf{43.62}$^\star$  & 25.98 &  19.65  & 29.76 &  10.89 &  4.41  & 16.59  & 29.30 &   \underline{\textbf{59.82}} &  \textbf{54.56}$^\star$ & -15.91 & 0 & \textbf{3} \\
        \textbf{MiniGPT4}  & -11.39 & -29.48 & -22.40  &  \underline{\textbf{41.99}}  & -2.92 &  42.23 &   8.69 &   6.28  & 33.21 & -24.96  & -23.26 &   1.55 &  24.31 &  41.81 &  28.09 & -92.14 & 0 & \textbf{1} \\
        \textbf{mPLUG-Owl} &  -51.63 & -39.29 & -43.82 & -17.88 & -43.61 & -31.29 & -33.59 & -39.34 & -27.84 & -66.64 &  -56.56 & -47.32 & -31.18 &   9.72 &  -4.55 & -86.22 & 0 & 0\\
        \textbf{mPLUG-Owl2.1} &   4.02 & -24.51 & -13.88 &  -5.57 &  -9.34 &  34.21 &  15.04 &  14.65 &  22.82 & -13.04 &   1.67 &   9.54 &  20.66  & 42.36 &  40.18 & -36.67 & 0 & 0 \\
        \textbf{AdapterV2} &   -6.59 & -46.57 & -55.69 &  \textbf{41.21}$^\star$ &  11.45 &   0.09 &  -8.55  & -9.48  & 20.20 &  -51.28 & -34.77  & -8.92 &  10.58 &   4.85 &  16.99 & -85.20 & 0 & \textbf{1} \\
        \textbf{VPGTrans}  & -43.92 & -62.55 & -53.01 &  12.25 & -13.51 &   8.77 & -20.50 &  -33.39 &  -4.39 & -47.00 & -47.45 & -38.11 & -14.56 & -23.58 &  24.73 & -95.06 & 0 & 0\\
        \textbf{MultiGPT}  & -68.31 & -71.87 & -65.42 & -52.14 & -57.55 & -51.68 & -48.83 & -55.39 & -45.53 & -82.97 & -56.97 & -52.24 & -38.90 &  -18.68 & -21.09 & -95.15 & 0 & 0\\
        \textbf{LaVIN-7B} & -26.60 &  -55.69 & -57.57 & -43.74 & -54.21 & -54.99 & -32.75  &-46.14 & -36.92 & -68.54  & -61.57 & -57.54 & -32.05 &   7.00  &   27.58 & -87.28 & 0 & 0\\
        \textbf{LaVIN-13B}  &  -23.10 &  -41.64 & -51.60 &  -32.30 &  -57.07 & -31.38  &-14.52 & -33.05 &  -9.08 & -46.96 &  -57.00 &   -45.79 & -21.40 & -4.87 &  -9.61  &-74.77 & 0 & 0\\
        \textbf{Lynx} &  -61.34 & -68.47 & -71.54 &  -0.42 & -16.14 & -22.35 & -31.13 & -67.05 & -14.45 & -81.81 & -34.28 & -17.50  &   0.44 &  -9.58 & -27.47 & -80.63 & 0 & 0\\
        \textbf{Fuyu-8B} &  -38.57 & -28.36 & -39.52 & -51.43 & -47.75 &  16.44 &   6.51  &-42.37 &  12.75 &   4.92 &  12.95 &   1.90  & -22.48 &  39.66  & 36.38 & -97.33 & 0 & 0\\
        \textbf{LaVIT} &    -35.15 & -33.32 & -44.78 &   5.81 & -12.38 &  -9.29 & -12.69 & -17.83 &  -2.38 & -63.04 & -37.07 & -13.85 &  -7.72 &  21.83 &   5.26 & -98.99 & 0 & 0\\
        \textbf{LLaMA-3.2-11B-Vision} & 17.78 &  61.75 &  54.90 &   -5.58 &  27.47  &-11.14 &   4.00 &    -0.50 &  -13.49 &  59.37 & 51.33 &  12.86 & -17.12 & -30.99 & -44.78 & \textbf{\textcolor[RGB]{34,139,34}{132.87}} & \textbf{1} & \textbf{1}\\
        \textbf{MiniCPM-V2.6} & \underline{\textbf{65.91}} & \underline{\textbf{108.09}}  & \textbf{95.10}$^\star$ &   11.02 &  \underline{\textbf{50.05}} &  24.88 &  20.67 &  48.38 & -25.38 &  63.65 &  60.17 &  \underline{\textbf{40.08}}  & 10.27 &  -3.69 & -31.43 & 112.60 & 0 & \textbf{6} \\
        \textbf{BLIP-2}  & 33.45 & -20.81 &  -4.40 &   33.24 &   7.23 &  \underline{\textbf{48.37}} &  \textbf{\textcolor[RGB]{34,139,34}{32.10}} &   44.64 &  \textbf{\textcolor[RGB]{34,139,34}{55.24}}  & 36.48 &   5.64 &  27.44 &  \textbf{33.17}$^\star$ &  53.54 &  \underline{\textbf{83.36}} &  67.07 & \textbf{2} & \textbf{5}  \\
        \textbf{InstructBLIP} & 37.34 & -21.23 &  -8.80 &   \textbf{\textcolor[RGB]{34,139,34}{42.63}} &  12.41 &  \textbf{\textcolor[RGB]{34,139,34}{55.90}} &  \underline{\textbf{31.24}} &  38.49 &  \underline{\textbf{53.27}} &  33.86 & 2.30 &   27.47 &  \textbf{\textcolor[RGB]{34,139,34}{37.10}} &   \textbf{54.82}$^\star$ &  \textbf{\textcolor[RGB]{34,139,34}{86.55}} &  75.62 & \textbf{\textcolor[RGB]{34,139,34}{4}} & \textbf{\textcolor[RGB]{34,139,34}{7}} \\
        \textbf{GLM-4V-9B} &  \textbf{\textcolor[RGB]{34,139,34}{72.26}} & \textbf{105.34}$^\star$ &  \underline{\textbf{95.99}} &   4.41 &  43.21 &  -4.55 &  13.65 &  39.65 & -21.05 &  62.74 & \textbf{\textcolor[RGB]{34,139,34}{64.72}}  & 36.75  &-10.50  & -22.38 & -25.30 &  122.47 & \textbf{2} & \textbf{4} \\
        \textbf{InternVL2.5-8B} &  \textbf{62.12}$^\star$  & 95.95 &  80.16 &   2.47  & 44.40 &   17.59 & 14.42 &  46.45 &  -0.46 &  \textbf{83.17}$^\star$ &   59.67  & \textbf{38.17}$^\star$  & 14.49 &   6.08 & -32.77  &111.1 & 0 & \textbf{3} \\
        \textbf{DeepSeek-VL2-tiny} & 22.98 & \textbf{\textcolor[RGB]{34,139,34}{113.17}} &  \textbf{\textcolor[RGB]{34,139,34}{99.40}}  &  -8.99 &  23.03 &  -4.78 &  17.68 & -28.24 & -32.96 &  48.85 &  52.23 &  13.51 &   3.69 & -93.03 & -35.12 &  63.04 & \textbf{2} & \textbf{2} \\
        \textbf{DeepSeek-VL2-small} &  20.95 &  48.05 &  40.61 & -21.39 &  25.14 & -32.48 &  -0.43 &  23.19 & -21.50 &   42.03 & 44.40 &   14.35 &  -0.74 & -93.87 & -44.01 &  36.85 & 0 & 0\\
        \textbf{DeepSeek-VL2} &  5.64 &  93.44 &  67.98 &  -2.47 &  42.39 &   7.92 &  17.57 &  16.49 & -22.74 &  57.75 & 40.14 &  19.65 &  -0.81 & -32.36 & -34.57 &  78.20 &  0 & 0\\ 
        \textbf{Qwen-VL-Chat} &  2.50 &  -27.45 &  -9.92 &  16.52 &  -0.21 &  38.59 &   5.99 &   5.20 &    9.87 &  15.57 & 14.69 &  23.83 &  15.74 &  37.05 &  53.65 & -21.78 &  0 & 0\\
        \textbf{Qwen2-VL-7B} & 45.94 &  56.13 &  61.85 &   7.08 &  40.20 &   -3.61 &  14.69 &  \underline{\textbf{56.41}}  &  4.45 &  77.74 &  59.60 &   \textbf{\textcolor[RGB]{34,139,34}{44.92}}  &  8.53  &-12.44 & -25.40 &  \textbf{\textcolor[RGB]{34,139,34}{132.87}} & \textbf{2} & \textbf{3} \\
        \textbf{Qwen2-VL-72B-Instruct} &   27.47 &  51.71 &  51.85  &  5.39  & \textbf{\textcolor[RGB]{34,139,34}{50.20}} &   42.12  & 18.23 &  \textbf{\textcolor[RGB]{34,139,34}{68.24}} &  16.94 &  \textbf{\textcolor[RGB]{34,139,34}{95.60}} & 42.69 &  34.13 &  26.97 &  35.99 & -26.68 & 120.77 & \textbf{3} & \textbf{3} \\
        \textbf{Qwen2.5-VL-3B-Instruct} &  39.48 &  53.59 &  42.88  &  6.55  & 44.73 &  11.91 &  10.11  & \textbf{48.67}$^\star$ &   4.13  & 80.97 &  \textbf{61.76}$^\star$ &  23.03 &  -9.13 & -36.63 & -21.74 & 118.28 & 0 & \textbf{2} \\
        \textbf{Qwen2.5-VL-7B-Instruct} &  56.65 &  81.83  & 79.81 &   9.79 &  \textbf{48.28}$^\star$ &  -0.85   & \textbf{26.35}$^\star$ &  47.72  &  8.72 &  \underline{\textbf{87.90}} &   \underline{\textbf{64.21}}  & 32.46 &   3.18  &  8.32 & -26.08 & \textbf{126.24}$^\star$ & 0 & \textbf{5} \\
       %  \textbf{\textcolor[RGB]{34,139,34}{} 绿色
       % \underline{\textbf{}}
       % \textbf{}$^\star$
        \bottomrule[1pt]
        \end{tabular}
        }
        
        \label{Table_model_mean_relative_gain}
        \end{center}
        % \vspace{-1em}
        \end{table*}

        \begin{table*}[!thp] \small
        \begin{center}
        \renewcommand{\arraystretch}{1.3} 
        \caption{The mean relative gain, MRG$^{\mathcal{I}}$ ($\uparrow$), for different instructions (\# 1, ..., \# 10) across  models.
        }
        \resizebox{1\textwidth}{!}{
        \begin{tabular}{
        p{1.6cm}< \centering| 
        p{1.3cm}< \centering p{1.3cm}< \centering p{1cm}< \centering |
        p{1.3cm}< \centering p{1.4cm}< \centering p{1.1cm}< \centering| 
        p{1.2cm}< \centering p{1.2cm}< \centering p{1.4cm}< \centering  p{1.4cm}< \centering| 
        p{1cm}< \centering p{1cm}< \centering p{1cm}< \centering|
        p{1cm}< \centering| p{1cm}< \centering| p{1cm}< \centering|
        p{0.5cm}< \centering p{0.6cm}< \centering
        }
        \toprule[1pt]
        \multirow{2}{*}{\textbf{Instructions}} 
        & \textbf{ScienceQA} & \textbf{PuzzleVQA} & \textbf{MMMU}
        & \textbf{MVSA-S} & \textbf{MVSA-M} & \textbf{TumEmo} &  \textbf{MOSI-2} & \textbf{MOSI-7} & \textbf{MOSEI-2} & \textbf{MOSEI-7} 
        & \textbf{Twitter15} & \textbf{Twitter17} &   \textbf{MASAD} & \textbf{Hate} & \textbf{Sarcasm} & \textbf{MNRE} & \textbf{Wins1} &\textbf{Wins3} \\
        \cline{2-19}
        & \multicolumn{17}{c}{\textbf{LLMs}} \\
        \cline{1-19}
        \textbf{\# 1} & -34.80 & - & - & \textbf{\textcolor[RGB]{34,139,34}{45.82}} &  \textbf{\textcolor[RGB]{34,139,34}{38.87}} &  \textbf{21.30}$^{\star}$ &   \textbf{\underline{23.31}} &  \textbf{\underline{33.34}} &  \textbf{\underline{24.02}} &  \textbf{\underline{38.88}} &  11.27 &  \textbf{\underline{26.92}} & \textbf{30.55}$^{\star}$ &  18.89 &   \textbf{\underline{28.83}} &  \textbf{26.64}$^{\star}$ & \textbf{2} & \textbf{\textcolor[RGB]{34,139,34}{11}} \\
        \textbf{\# 2} & 10.86 & - & - &  \textbf{\underline{39.77}} &  \textbf{\underline{32.15}} &  \textbf{\textcolor[RGB]{34,139,34}{22.85}} &  \textbf{\textcolor[RGB]{34,139,34}{24.99}} &  \textbf{\textcolor[RGB]{34,139,34}{34.09}} &  \textbf{\textcolor[RGB]{34,139,34}{28.33}} &  \textbf{\textcolor[RGB]{34,139,34}{42.55}} &  15.34 &  26.06 & \textbf{\textcolor[RGB]{34,139,34}{40.88}} &  \textbf{20.59}$^{\star}$ &  \textbf{17.18}$^{\star}$ & -13.15 & \textbf{\textcolor[RGB]{34,139,34}{6}} & \textbf{10} \\
        \textbf{\# 3} & 5.35 & - & - &  -31.52 & -20.19 & -27.74 &   7.14 &   1.47 &  -8.40 &  -14.86 &  \textbf{\underline{47.78}} &  20.12 & -9.04 &   4.36 & -14.16 &  18.94 & 0 & \textbf{1} \\
        \textbf{\# 4} & -31.96 & - & - & -34.92 & -30.23 &  -9.90 &  -29.31 & -16.24 & -25.12 & -14.93 & -52.40 &  -54.08 &  -55.31 &   2.22 &  \textbf{\textcolor[RGB]{34,139,34}{30.90}} &  -45.35 & \textbf{1} & \textbf{1}\\
        \textbf{\# 5} &   \textbf{\underline{40.33}} & - & - &   3.15 &  14.05 &  13.52 &  \textbf{20.05}$^{\star}$  &  \textbf{19.21}$^{\star}$ &   0.13 &  \textbf{14.49}$^{\star}$  & \textbf{\textcolor[RGB]{34,139,34}{48.73}} &  \textbf{\textcolor[RGB]{34,139,34}{32.27}} &   12.54 &  \textbf{\textcolor[RGB]{34,139,34}{21.87}} &   2.24 &  \textbf{\textcolor[RGB]{34,139,34}{97.96}} & \textbf{4} & \textbf{8}\\
        \textbf{\# 6} & \textbf{27.15}$^{\star}$ & - & - &  8.38 &  10.18 &  19.21 & -12.44 &  -4.35 &  -5.02 &   4.19 &   3.59 &   6.33 &  14.04 & -32.51 & -12.14 & -50.15 & 0 & 1\\
        \textbf{\# 7} & \textbf{\textcolor[RGB]{34,139,34}{41.61}} & - & - & \textbf{20.04}$^{\star}$ &  \textbf{21.21}$^{\star}$ &  \textbf{\underline{21.77}} &   5.16 & -26.04 &  \textbf{17.59}$^{\star}$ & -27.95 &  \textbf{20.66}$^{\star}$  & \textbf{26.78}$^{\star}$ &  \textbf{\underline{39.83}} &   \textbf{\underline{21.71}} &  12.38 &   \textbf{\underline{61.19}} & \textbf{1} & \textbf{10} \\
        \textbf{\# 8} & -37.11 & - & - &  4.09 &   4.95 &  -7.54 &  -8.42 &  -4.17 &   4.90 &   -7.24 & -42.70 &  -38.90 & -36.50 &   -1.31 & -16.69 & -47.31 & 0 & 0\\
        \textbf{\# 9} &  4.22 & - & - & -31.77 & -38.78 & -19.21 &   2.24 &   0.36 & -10.30 &   -0.44  &-14.47 &  -7.95 & -0.67 & -41.00 &   -22.52 & -26.07 & 0 & 0 \\
        \textbf{\# 10} & -25.66 & - & - & -23.04 & -32.21 & -34.26 & -32.72 & -37.66 & -26.13 & -34.68 & -37.82 & -37.55 & -36.32 & -14.82 & -26.02  &-22.72 & 0 & 0 \\
        \midrule[1pt]
        & \multicolumn{17}{c}{\textbf{MLLMs}} \\
        \cline{1-19}
        % \midrule[1pt]
        \textbf{\# 1} & -19.75 & -13.23 & -10.30 & \textbf{10.31}$^\star$ & 5.53 & \textbf{\textcolor[RGB]{34,139,34}{17.92}} & \underline{\textbf{15.73}} & \underline{\textbf{19.69}} & \textbf{6.83}$^\star$ & \underline{\textbf{16.37}} & 1.53 & 7.05 & 0.27 & -2.78 & \textbf{6.55}$^\star$ & 1.24 & \textbf{1} & \textbf{7}\\  
        \textbf{\# 2} & 7.72 & 7.76 & \textbf{7.30}$^\star$ & \underline{\textbf{16.74}} & \textbf{\textcolor[RGB]{34,139,34}{18.83}} & \underline{\textbf{15.26}} & \textbf{14.98}$^\star$ & \textbf{\textcolor[RGB]{34,139,34}{31.33}} & \underline{\textbf{12.37}} & \textbf{\textcolor[RGB]{34,139,34}{24.94}} & \underline{\textbf{10.90}} & \textbf{\textcolor[RGB]{34,139,34}{15.13}} & \underline{\textbf{12.37}} & \textbf{8.92}$^\star$ & \textbf{\textcolor[RGB]{34,139,34}{10.51}} & 0.66 & \textbf{\textcolor[RGB]{34,139,34}{5}} & \textbf{\textcolor[RGB]{34,139,34}{13}} \\
        \textbf{\# 3} & \textbf{14.11}$^\star$ & \underline{\textbf{14.51}} & 6.22 & -10.21 & -6.26 & -7.74 & -4.13 & -16.42 & -1.20 & -8.80 & 2.70 & -2.57 & -4.43 & 5.63 & -7.63 & \textbf{\textcolor[RGB]{34,139,34}{20.19}} & \textbf{1} & \textbf{3} \\
        \textbf{\# 4} & -13.44 & -15.88 & -13.86 & -5.53 & -4.12 & -4.56 & -6.68 & -1.77 & -6.18 & -1.05 & -7.47 & -6.63 & -7.68 & -6.64 & 3.49 & -17.83  & 0 & 0 \\
        \textbf{\# 5} & \textbf{\textcolor[RGB]{34,139,34}{29.24}} & \textbf{\textcolor[RGB]{34,139,34}{15.86}} & \textbf{\textcolor[RGB]{34,139,34}{17.00}} & -10.11 & -7.27 & -10.12 & 4.67 & -7.49 & 3.68 & -8.20 & \textbf{\textcolor[RGB]{34,139,34}{11.85}} & 4.87 & 0.52 & \underline{\textbf{18.63}} & 2.11 & \underline{\textbf{15.51}} & \textbf{4} & \textbf{6} \\
        \textbf{\# 6} & -6.04 & -4.25 & 0.98 & 6.57 & \textbf{10.89}$^\star$ & 2.63 & -7.77 & 0.84 & -0.73 & \textbf{7.52}$^\star$ & \textbf{9.76}$^\star$ & \underline{\textbf{8.33}} & \textbf{9.06}$^\star$ & -7.43 & 1.40 & 0.31 & 0 & \textbf{5} \\
        \textbf{\# 7} & \underline{\textbf{19.13}} & \textbf{8.57}$^\star$ & \underline{\textbf{11.04}} & \textbf{\textcolor[RGB]{34,139,34}{17.00}} & \underline{\textbf{11.39}} & \textbf{5.90}$^\star$ & \textbf{\textcolor[RGB]{34,139,34}{17.61}} & -10.83 & \textbf{\textcolor[RGB]{34,139,34}{22.93}} & -3.01 & 7.20 & \textbf{8.25}$^\star$ & \textbf{\textcolor[RGB]{34,139,34}{18.86}} & \textbf{\textcolor[RGB]{34,139,34}{22.80}} & \underline{\textbf{10.44}} & \textbf{2.10}$^\star$ & \textbf{\textcolor[RGB]{34,139,34}{5}} & \textbf{\textcolor[RGB]{34,139,34}{13}} \\
        \textbf{\# 8} & -18.06 & -9.34 & -10.81 & -11.36 & -12.90 & -5.47 & -19.73 & -10.53 & -22.10 & -18.24 & -20.04 & -20.28 & -19.21 & -17.77 & -11.29 & -12.11 & 0 & 0 \\
        \textbf{\# 9} & 2.93 & 5.79 & 1.51 & -1.32 & -2.14 & -3.02 & 0.05 & \textbf{8.84}$^\star$ & 0.26 & 5.07 & -3.63 & -3.13 & -1.63 & -11.52 & -8.64 & -5.73 & 0 & \textbf{1} \\
        \textbf{\# 10} & -15.83 & -9.80 & -9.09 & -12.08 & -13.96 & -10.81 & -14.73 & -13.66 & -15.86 & -14.61 & -12.80 & -11.02 & -8.13 & -9.84 & -6.94 & -4.35 & 0 & 0 \\
        \bottomrule[1pt]
        \end{tabular}
        }
        \label{Table_instruction_mean_relative_gain}
        \end{center}
        % \vspace{-1em}
        \end{table*}

        \begin{table*}[h] \small
        \begin{center}
        \renewcommand{\arraystretch}{1.2} 
        \caption{The stability, $S^{\mathcal{M^{'}}}$ ($\downarrow$), of various models with excellent performance across instructions. 
        }
        % \vspace{-1em}
        \resizebox{1\textwidth}{!}{
        \begin{tabular}{
        p{3.8cm}< \centering| 
        p{1.3cm}< \centering p{1.3cm}< \centering p{1cm}< \centering |
        p{1.3cm}< \centering p{1.4cm}< \centering p{1.1cm}< \centering| 
        p{1.2cm}< \centering p{1.2cm}< \centering p{1.4cm}< \centering  p{1.4cm}< \centering| 
        p{1cm}< \centering p{1cm}< \centering p{1cm}< \centering|
        p{1cm}< \centering| p{1cm}< \centering| p{1cm}< \centering|
        p{0.6cm}< \centering
        }
        \toprule[1pt]
        \textbf{Models}
        & \textbf{ScienceQA} & \textbf{PuzzleVQA} & \textbf{MMMU}
        & \textbf{MVSA-S} & \textbf{MVSA-M} & \textbf{TumEmo} &  \textbf{MOSI-2} & \textbf{MOSI-7} & \textbf{MOSEI-2} & \textbf{MOSEI-7} 
        & \textbf{Twitter15} & \textbf{Twitter17} & \textbf{MASAD} 
        & \textbf{Hate} 
        & \textbf{Sarcasm} 
        & \textbf{MNRE}
        & \textbf{Wins1} \\
        \midrule[1pt]
        \textbf{\textcolor{red}{ChatGPT}} & 4.36 & - & - & 13.44  & 8.37 &  3.61 &  1.56 &  2.17  & 5.90  &  3.57 &  2.53  & 3.20  &  6.32  & 6.90  & 9.75  & 6.92 & -  \\
        \textbf{LLaMA3.1-8B-Instruct} & 17.61 & - & - & 18.16 & 16.78 & 12.30 &  17.11 & 11.86 & 17.36 & 10.47 &  8.35 &  8.59 & 11.34 &  8.20  & 7.99 &  5.08 & - \\
        \textbf{Flan-T5-XXL} & \textbf{\textcolor[RGB]{34,139,34}{0.57}} & - & - & 10.84 &  8.99 &  4.62 &  0.73 &  3.73 &  \textbf{\textcolor[RGB]{34,139,34}{0.38}} &  2.97 &  9.63 &  3.09  & \textbf{\textcolor[RGB]{34,139,34}{0.52}} &  0.93 &  1.75 &  2.19 & -  \\
        \midrule[1pt]
        \textbf{\textcolor{red}{Gemini-V}} &  15.69 & \textbf{\textcolor[RGB]{34,139,34}{0.39}} & 4.97 & 11.35 & 8.71 & 6.59 & 8.41 & 7.40 & 3.91 & 6.46 & 7.83 & 5.53 & 7.38 & 10.24 & 4.93 & 5.81 & \textbf{1} \\
        \textbf{LLaVA-v1.6-13B} & 4.70 & 0.51 & 1.51 & 5.17 & 5.48 & 2.61 & 1.85 & 2.59 & 5.79 & 3.07 & 5.95 & 2.86 & 4.42 & 1.43 & 2.76 & 5.74 & 0 \\
        \textbf{MiniGPT4} & 6.71 & 2.56 & 2.35 & 4.99 & 13.06 & 2.48 & 9.17 & 6.62 & 8.02 & 7.10 & 3.93 & 4.23 & 3.91 & 1.65 & 5.95 & \textbf{\textcolor[RGB]{34,139,34}{0.69}} & \textbf{1} \\
        \textbf{AdapterV2} & 2.64 & 7.78 & 7.88 & 10.65 & 8.46 & 6.68 & 23.57 & 11.08 & 21.00 & 7.26 & 7.14 & 12.06 & 20.19 & 15.25 & 6.61 & 2.13 & 0 \\
        \textbf{LLaMA-3.2-11B-Vision} & 15.55 & 6.72 & 7.87 & 6.87 & 11.39 & 5.39 & 7.86 & 4.34 & 6.02 & 4.25 & 2.36 & 10.04 & 6.11 & 4.75 & 4.43 & 2.39 & 0 \\
        \textbf{MiniCPM-V2.6} & 2.65 & 1.28 & 1.28 & \textbf{\textcolor[RGB]{34,139,34}{0.85}} & \textbf{\textcolor[RGB]{34,139,34}{0.65}} & 3.45 & 1.96 & 3.17 & 2.02 & 1.02 & 0.78 & 1.07 & 3.42 & 3.03 & 0.87 & 1.44 & \textbf{2} \\
        \textbf{BLIP-2}  & 5.73 & 0.99 & 1.29 & 5.46 & 7.86 & 3.05 & \textbf{\textcolor[RGB]{34,139,34}{0.31}} & 3.62 & \textbf{\textcolor[RGB]{34,139,34}{0.42}} & 3.43 & 9.04 & 3.55 & \textbf{\textcolor[RGB]{34,139,34}{0.53}} & 0.69 & 1.52 & 1.89 & \textbf{3} \\
        \textbf{InstructBLIP} & \textbf{\textcolor[RGB]{34,139,34}{0.73}} & 1.46 & \textbf{\textcolor[RGB]{34,139,34}{1.07}} & 4.15 & 4.44 & \textbf{\textcolor[RGB]{34,139,34}{0.62}} & 0.62 & 5.02 & 0.51 & 4.24 & 7.02 & 3.29 & 0.92 & \textbf{\textcolor[RGB]{34,139,34}{0.65}} & 1.27 & 3.90 & \textbf{\textcolor[RGB]{34,139,34}{4}} \\
        \textbf{GLM-4V-9B} & 3.46 & 3.15 & 1.60 & 3.32 & 3.36 & 8.62 & 4.01 & 5.91 & 4.39 & 1.69 & 0.45 & 1.57 & 6.41 & 2.52 & \textbf{\textcolor[RGB]{34,139,34}{0.51}} & 1.69 & \textbf{1} \\
        \textbf{InterVL2.5-8B} & 4.63 & 1.98 & 1.63 & 1.27 & 1.08 & 1.65 & 2.70 & 1.90 & 2.72 & 1.08 & 0.51 & 0.96 & 0.95 & 3.07 & 0.91 & 3.07 & 0 \\
        \textbf{DeepSeekVL2-tiny} & 5.54 & 2.00 & 1.40 & 2.80 & 8.52 & 1.86 & 3.07 & 3.35 & 6.06 & 2.30 & 2.37 & 1.34 & 6.89 & 5.14 & 0.74 & 1.95 & 0 \\
        \textbf{Qwen2-VL-7B-Instruct} &  3.15 & 5.36 & 6.44 & 0.87 & 7.67 & 6.18 & 8.16 & 1.41 & 4.95 & 2.26 & 0.50 & \textbf{\textcolor[RGB]{34,139,34}{0.89}} & 2.50 & 2.29 & 1.45 & 2.39 & \textbf{1} \\
        \textbf{Qwen2-VL-72B-Instruct} & 2.31 & 2.74 & 2.49 & 1.39 & 1.14 & 1.22 & 4.25 & \textbf{\textcolor[RGB]{34,139,34}{1.18}} & 2.68 & \textbf{\textcolor[RGB]{34,139,34}{0.49}} & 3.67 & 1.01 & 3.75 & 1.43 & 1.34 & 2.40 & \textbf{2} \\
        \textbf{Qwen2.5-VL-3B-Instruct} & 2.68 & 9.42 & 8.70 & 1.60 & 1.56 & 2.41 & 3.50 & 7.98 & 7.32 & 2.14 & 0.56 & 6.51 & 5.02 & 2.91 & 0.56 & 1.65 &  0\\
        \textbf{Qwen2.5-VL-7B-Instruct} & 2.21 & 4.52 & 2.00 & 1.82 & 2.06 & 5.27 & 0.86 & 7.51 & 6.37 & 3.26 & \textbf{\textcolor[RGB]{34,139,34}{0.31}} & 4.80 & 4.86 & 2.53 & 1.44 & 1.80 & \textbf{1} \\
       \bottomrule[1pt]
        \end{tabular}
        }
        \label{stability of various models}
        \end{center}
        % \vspace{-1em}
        \end{table*}
        % \textbf{\textcolor[RGB]{34,139,34}{} 绿色
       % \underline{\textbf{}}
       % \textbf{}$^\star$
        
        \subsection{Mean Relative Gain}
        
        While the best performance metric offers insight into the upper limit of capabilities for
        each model, a comprehensive assessment on the each full dataset requires a broader perspective.
        \subsubsection{Mean Relative Gain of Models}
        To assess the comprehensive performance of each model, we calculate the mean relative gain for each model across all instructions on each dataset, as detailed in Eq. \ref{Eq_model_MRG}. The results are tabulated in Table \ref{Table_model_mean_relative_gain}. The metrics ``\textbf{Wins1}'' and ``\textbf{Wins3}'' denote the number of tasks where a model or an instruction achieves the highest and top-three aggregated performances, respectively. Similar to the best model performance metric, models equipped with the Flan-T5-XXL backbone, such as InstructBLIP, BLIP-2, and Flan-T5-XXL itself, exhibit strong performance. The latest models, notably Qwen2.5-VL and MiciCPM-V2.6, demonstrate superior performance across all instructions, particularly in VQAMC tasks. However, their strong performance should be interpreted with caution, as they may have been pre-trained on the ScienceQA dataset, potentially introducing data leakage that gives them an advantage over other models.
        Closed-source models also show excellent results.
        \subsubsection{Mean Relative Gain of Instructions}
        We continue to evaluate the overall efficacy of each instruction by calculating the mean relative gain (MRG$^{\mathcal{I}}$) across all models, as detailed in Eq. \ref{Eq_Instruction_MRG}. The outcomes are presented in Table \ref{Table_instruction_mean_relative_gain}. Due to the different evaluation datasets for LLMs and MLLMs—with LLMs excluding PuzzleVQA and MMMU and MLLMs including these datasets—we compute MRG$^\mathcal{I}$ separately for each group.
        Our analysis reveals that Instructions \# 1, \# 2, \# 5, and \# 7 generally perform better, with their detailed designs illustrated in Figures \ref{Figure_1_PuzzleVQA_different_instructions} and \ref{Figure_1_MSD_Multimodal_Instruction}. Notably, Instructions \# 5 and \# 7, which evolve from simple ``\textbf{Question-Answer}'' format of \# 2, show superior performance. The effectiveness likely stems from the prior training of models on related QA tasks, where specific terms like ``Question'' and ``Answer'' enhance model responsiveness. Furthermore, the inclusion of option terms in Instructions \# 5 and \# 7 appears to influence performance on specific datasets positively.
        We find an interesting phenomenon, where models performed better on VQAMC tasks when given Instruction \# 5, which includes options, while Instruction \# 2, without options, yielded superior results for MSA tasks. This pattern may be attributed to the distinct nature of these tasks. In VQAMC, each question has unique answers, making the provided options valuable hints for response generation. Conversely, in MSA tasks, the label space remains constant across examples, rendering the inclusion of options less beneficial.
        
        \subsection{Stability}
        Stability is a key metric for assessing both models and instructions on each full dataset. Notably, some models display consistent performance across all instructions, resulting in high stability, albeit at lower performance levels. Consequently, our evaluations prioritize the stability of models and instructions that show consistently positive performance.
        % \vspace{-1em}
        \subsubsection{Stability of Models}
        \begin{table*}[t] \small
        \begin{center}
        \renewcommand{\arraystretch}{1.3} 
        \caption{The stability, $S^{\mathcal{I^{'}}}$ ($\downarrow$), of different instructions with excellent performance across models.
        }
        \resizebox{1\textwidth}{!}{
        \begin{tabular}{
        p{1.8cm}< \centering| 
        p{1.3cm}< \centering p{1.3cm}< \centering p{1cm}< \centering |
        p{1.3cm}< \centering p{1.4cm}< \centering p{1.1cm}< \centering| 
        p{1.2cm}< \centering p{1.2cm}< \centering p{1.4cm}< \centering  p{1.4cm}< \centering| 
        p{1cm}< \centering p{1cm}< \centering p{1cm}< \centering|
        p{1cm}< \centering| p{1.1cm}< \centering| p{1cm}< \centering|
        p{0.5cm}< \centering 
        }
        \toprule[1pt]
        \multirow{2}{*}{\textbf{Instructions}} 
        & \textbf{ScienceQA} & \textbf{PuzzleVQA} & \textbf{MMMU}
        & \textbf{MVSA-S} & \textbf{MVSA-M} & \textbf{TumEmo} &  \textbf{MOSI-2} & \textbf{MOSI-7} & \textbf{MOSEI-2} & \textbf{MOSEI-7} 
        & \textbf{Twitter15} & \textbf{Twitter17} &   \textbf{MASAD} & \textbf{Hate} & \textbf{Sarcasm} & \textbf{MNRE} & \textbf{Wins1} \\
        \cline{2-18}
        & \multicolumn{16}{c}{\textbf{LLMs}} \\
        \cline{1-18}
        \textbf{\# 1} & 21.96 & - & - & 8.81 &  9.59 &  8.28 &  \textbf{\textcolor[RGB]{34,139,34}{8.96}} &  \textbf{\textcolor[RGB]{34,139,34}{6.17}} &  8.15 & 11.56 & 11.65 & 7.83 &  8.36  & 8.48 & 6.89 & 13.48 & \textbf{2} \\
        \textbf{\# 2} & 15.33 & - & - &  \textbf{\textcolor[RGB]{34,139,34}{7.60}} & 12.38 &  \textbf{\textcolor[RGB]{34,139,34}{5.65}} &  9.81 &  6.28 &  \textbf{\textcolor[RGB]{34,139,34}{7.95}} & \textbf{\textcolor[RGB]{34,139,34}{10.63}} & \textbf{\textcolor[RGB]{34,139,34}{11.54}} &  8.16  & \textbf{\textcolor[RGB]{34,139,34}{3.50}}  &  7.32 &  11.93 & 16.01 & \textbf{\textcolor[RGB]{34,139,34}{6}} \\
        \textbf{\# 3} & 22.65 & - & - & 23.67 & 16.14 & 18.05 & 15.06 & 10.75 & 18.82 & 12.83  & 12.52 & 11.85 & 23.14 & 16.60 &  18.62 & 12.18 & 0 \\
        \textbf{\# 4} & 24.03 & - & - & 24.72 & 25.36 & 13.66 & 30.93 & 13.60 &  25.72 & 14.02 & 22.67 & 23.40  & 29.70 &  12.49 &  \textbf{\textcolor[RGB]{34,139,34}{6.04}} & 15.08 & \textbf{1} \\
        \textbf{\# 5} & 12.90 & - & - &  15.34 & \textbf{\textcolor[RGB]{34,139,34}{10.84}} & 10.94 & 11.59 &  6.99 & 17.77 & 11.65 & 13.19 &  \textbf{\textcolor[RGB]{34,139,34}{7.59}} & 15.26 &  \textbf{\textcolor[RGB]{34,139,34}{4.70}} &  11.93 & \textbf{\textcolor[RGB]{34,139,34}{10.74}} & \textbf{4} \\
        \textbf{\# 6} & \textbf{\textcolor[RGB]{34,139,34}{12.34}} & - & - & 13.11 & 16.27 &  9.09 & 23.09 & 11.79 & 21.25 & 14.06  &11.96 & 12.05 & 12.22 & 20.85 & 20.70 &  14.20 & \textbf{1} \\
        \textbf{\# 7} & 12.97 & - & - &  8.80 &  12.21 &  6.87 & 17.40 &  15.49 &  8.55 & 14.79 & 11.94 &  9.33 &  3.92 &  7.65 &  16.84 & 10.45 & 0 \\ 
        \midrule[1pt]
        & \multicolumn{16}{c}{\textbf{MLLMs}} \\
        \cline{1-18}
        % \midrule[1pt]
        \textbf{\# 1} & 26.44 & 21.76 & 21.12 & \textbf{\textcolor[RGB]{34,139,34}{11.42}} & 21.84 & \textbf{\textcolor[RGB]{34,139,34}{9.98}} & 10.13 & 11.65 & 15.90 & 18.39 & 22.75 & 15.58 & 17.65 & 16.83 & 16.20 & 17.10 & \textbf{2} \\
        \textbf{\# 2} & 19.90 & 19.94 & 19.80 & 13.02 & \textbf{\textcolor[RGB]{34,139,34}{19.86}} & 10.04 & \textbf{\textcolor[RGB]{34,139,34}{9.67}} & \textbf{\textcolor[RGB]{34,139,34}{9.93}} & 15.61 & \textbf{\textcolor[RGB]{34,139,34}{17.70}} & 21.43 & \textbf{\textcolor[RGB]{34,139,34}{12.96}} & 13.03 & 15.60 & 14.53 & 17.85 & \textbf{\textcolor[RGB]{34,139,34}{5}} \\
        \textbf{\# 3} & 22.72 & 19.23 & 18.59 & 20.02 & 26.46 & 14.18 & 19.27 & 10.81 & 18.39 & 19.55 & 22.62 & 15.94 & 19.16 & 16.98 & 18.37 & 17.32 & 0\\
        \textbf{\# 5} & 19.95 & 19.43 & \textbf{\textcolor[RGB]{34,139,34}{17.81}} & 18.88 & 26.35 & 14.82 & 16.92 & 11.17 & 16.29 & 19.46 & 22.47 & 15.52 & 17.99 & 16.32 & 18.35 & 18.36 & \textbf{1} \\
        \textbf{\# 6} & 22.23 & 21.47 & 20.88 & 14.18 & 21.09 & 11.78 & 24.19 & 13.28 & 20.48 & 19.16 & 22.59 & 15.05 & 15.97 & 17.10 & \textbf{\textcolor[RGB]{34,139,34}{14.30}} & \textbf{\textcolor[RGB]{34,139,34}{16.31}} & \textbf{2 } \\
        \textbf{\# 7} & \textbf{\textcolor[RGB]{34,139,34}{17.28}} & 18.84 & 17.95 & 14.29 & 22.20 & 10.89 & 10.95 & 12.54 & \textbf{\textcolor[RGB]{34,139,34}{14.04}} & 20.05 & \textbf{\textcolor[RGB]{34,139,34}{21.33}} & 13.65 & \textbf{\textcolor[RGB]{34,139,34}{11.12}} & \textbf{\textcolor[RGB]{34,139,34}{15.14}} & 15.56 & 17.70 & \textbf{\textcolor[RGB]{34,139,34}{5}} \\
        \textbf{\# 9} & 19.96 & \textbf{\textcolor[RGB]{34,139,34}{18.67}} & 18.85 & 17.23 & 24.42 & 12.68 & 19.95 & 11.91 & 21.02 & 18.86 & 23.90 & 16.29 & 17.81 & 17.96 & 15.88 & 16.60 & \textbf{1} \\
        \bottomrule[1pt]
        \end{tabular}
        }
        \label{Figure_stability_of_various_instruction}
        \end{center}
        % \vspace{-1em}
        \end{table*}

        We exclusively use Eq. \ref{Eq_model_stability} to calculate stability metrics for models demonstrating strong overall performance, such as ChatGPT, Flan-T5-XXL, Gemini-V, LLaMA-3.2-11B-Vision, Qwen2-VL, Qwen2.5-VL, MiniCPM-V2.6, BLIP2, InstructBLIP, and others. The results are presented in Table \ref{stability of various models}.
        Among open-source models, the Flan-T5 series exhibits minimal fluctuations across different instructions, indicating lower sensitivity to instruction variability. Compared to BLIP2, InstructBLIP enhances stability by incorporating instruction tuning, effectively reducing the model's sensitivity to varying instructions in multimodal reasoning tasks with vision-text contexts. This improvement highlights instruction tuning as a valuable strategy for enhancing model stability.
        Additionally, the recently released MiniCPM-V2.6 and Qwen2-VL-72B-Instruct combine exceptional performance with remarkably low sensitivity. This suggests that these models are specifically designed with instruction tuning and additional training in mind to optimize stability and performance.

        \subsubsection{Stability of Instructions}

        We assess the stability of instructions that achieve a top-three average relative gain on at least one dataset from LLMs and MLLMs using Eq. \ref{Eq_Instruction_stability}, as detailed in Table \ref{Figure_stability_of_various_instruction}. Notably, Instruction \# 2 exhibits the highest stability and also demonstrates strong aggregated performance. The superior stability likely stems from the prevalent training of models on the ``Question-Answer'' data format. The finding provides valuable insights for the design of future instructions.
        
        \subsection{Adaptability}
        Models are trained with distinct pre-training settings, leading to varying preferences towards different instructions. We quantify the adaptability using the Global Top-K Hit Ratio, as detailed in Eq. \ref{Eq_hit_ratio}. To thoroughly assess the adaptability of various models to different instructions across all full datasets, we have categorized all models into nine groups based on their language backbones. These groups include Closed-source, LLaMA LLMs, Other LLMs, MLLMs based on LLaMA, MLLMs based on Vicuna, MLLMs based on Flan-T5, LLaVA Series, LaVIT Series, DeepSeek Series, Qwen-VL Series, and MLLMs based on other language backbones. The comprehensive results, covering all datasets, are illustrated in Figure \ref{Figure_Adaptability_leida}. It's important to note that evaluations for LLMs exclude the AlgoPuzzleVQA and MMMU datasets, while MLLMs include the two datasets.
        
        \subsubsection{Total Adaptability}
        \begin{figure*}[t] %%
          \centering %?????????
          \includegraphics[width = 1\textwidth]{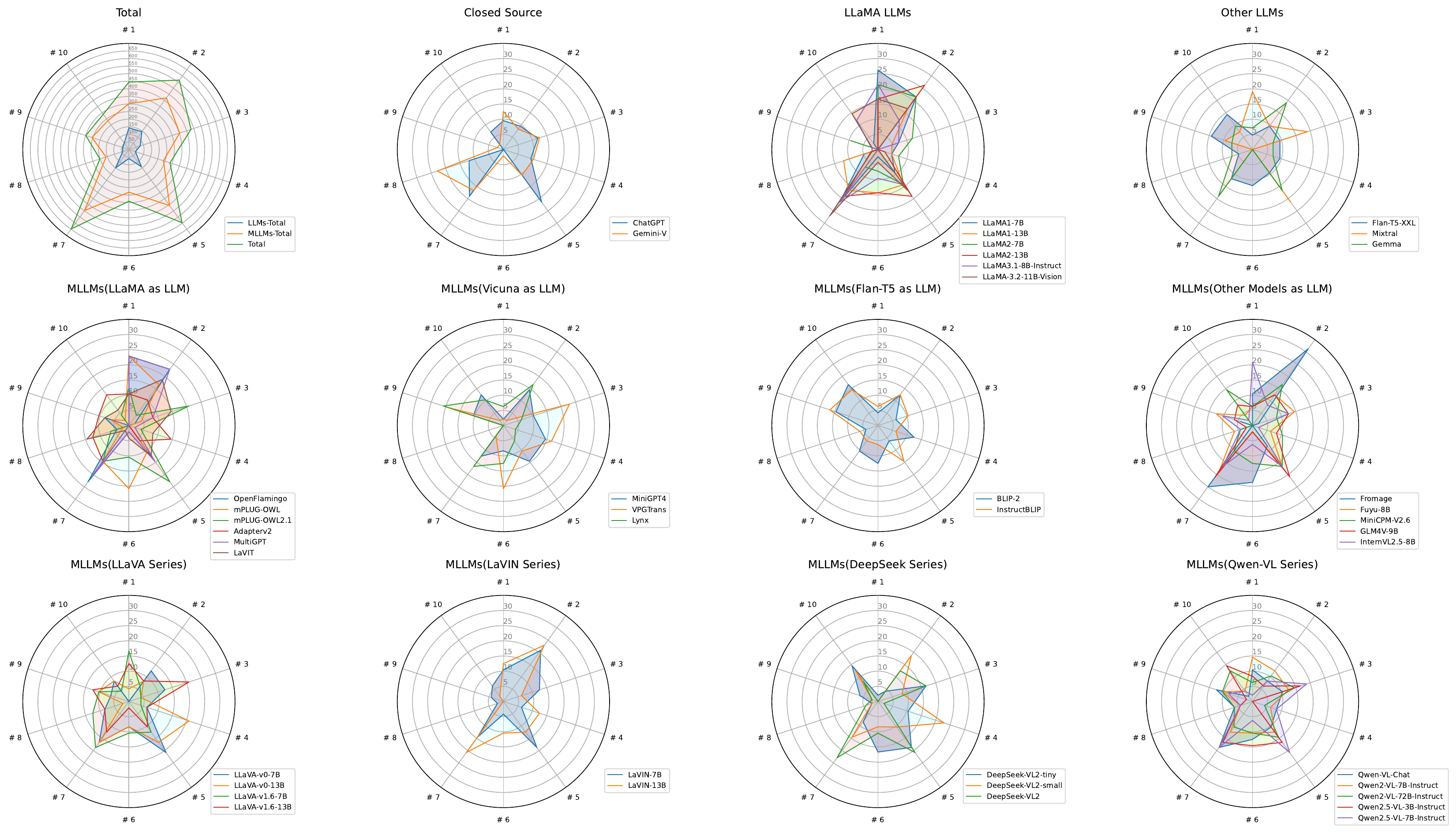
          }
          %\vpsace{-1em}
          \caption{The Top-K Hit Ratio, $GHR@K$ ($\uparrow$), for various instructions (\# 1, ..., \# 10) across different models on all datasets. The ``\textbf{Total}'' represents the sum of Global Top-K Hit Ratio scores across all datasets for each model.
           }
          \label{Figure_Adaptability_leida} 
          %\vpsace{-1.5em}
        \end{figure*}
        Overall, instructions \# 2 and \# 7 demonstrated the highest adaptability among all models as the ``\textbf{Total}'' subplot shown in Figure \ref{Figure_Adaptability_leida}, indicating that these instructions should be prioritized in the design of future models.
        % \vspace{-1em}
        \subsubsection{Adaptability for Different Models}
        We advert that different models tend to perform better with specific instructions. 
        (1) \textbf{\textcolor{red}{Closed-source} models}: Closed-source models exhibit their poorest performance on Instructions \# 6 and \# 9, suggesting that specific trigger words such as ``AI:'', ``Human:'', and ``\#\#\# Input'' significantly impair their performance.
        (2) \textbf{Same models with different version}:
        The LLaMA model series, including LLaMA1-7B, LLaMA1-13B, LLaMA2-7B, LLaMA2-13B, and LLaMA-3.2-11B-Vision, consistently demonstrates strong performance on Instruction \# 2, indicating similar preferences across different models. Similarly, the LaVIN model branch excels with Instructions \# 2 and \# 7. The Qwen-VL series achieves better performance on Instructions \# 3, \# 5, and \# 7.
        In contrast, the LLaVA model branch and the DeepSeek series exhibit more varied preferences, as models of different sizes optimize performance using different instructions. This variation may be attributed to differences in training methodologies applied at different scales.
        (3) \textbf{Various backbones of LLMs}:
        We further categorize additional models into four groups based on large language model backbones. Related groups are listed that models using LLaMA, Vicuna, Flan-T5, and others as their LLMs.
        We note that models based on different backbones exhibit varied performance across diverse instructions. For instance, most models utilizing the LLaMA backbone excel on Instructions \# 1, \# 2, and \# 7. However, certain models, such as OpenFlamingo and mPLUG-OWL, struggle with specific instructions, achieving an adaptability score of zero on \# 4, among others.
        Models with Vicuna backbones generally underperform on Instructions \# 8 and \# 1. Conversely, the Flan-T5 series, which includes Flan-T5, BLIP-2, and InstructBLIP, shows consistent adaptability trends, performing well on Instructions \# 9 and \# 10 but less so on Instructions \# 1 and \# 8.
        Models leveraging other LLM backbones display optimal performance on a range of instructions, with Fromage, in particular, experiencing significant performance variability across different instructions.
        
        Our observation reaffirms that the training of each model involves carefully designed instructions, including specific trigger words. Consequently, to achieve the best performance, it is essential to tailor instructions for each model based on the prompts used during its training.
        Furthermore, \textsc{MM-InstructEval} offers crucial guidance and direction for crafting instructions that maximize the performance of both existing and emerging models.

	\section{Conclusion}
	\label{Conclusion}
        In this work, we present MM-InstructEval, a comprehensive evaluation framework that assesses 45 models (including 36 MLLMs) across 16 diverse multimodal reasoning datasets using 10 different instruction formats. Our systematic evaluation framework introduces multiple innovative metrics, the Best Performance metric, Mean Relative Gain metric, Stability metric, and Adaptability metric, enabling a thorough and multi-faceted assessment of model capabilities.
        Through our extensive evaluation, we uncover several significant findings that provide valuable insights.
        First, newer models achieve comparable or superior performance with substantially fewer parameters compared to their predecessors. This finding indicates significant advances in model architecture design and training methodologies, suggesting a shift toward more efficient and practical implementations of MLLMs.
        Second, we reveal that ``Question-Answer'' style instructions generally lead to better performance across models. However, we also observe notable variations in instruction preferences across different models, highlighting the importance of careful prompt engineering in practical applications.
        Third, our evaluation demonstrates the superior capabilities of advanced closed-source models like GPT-4V and GLM-4V-plus, particularly in handling challenging datasets and complex reasoning tasks. This performance gap provides valuable benchmarks for ongoing development in open-source alternatives.
        
        Research on multimodal large language models continues to advance at a rapid pace. Our comprehensive evaluation framework and findings provide a robust foundation for understanding these systems' capabilities and limitations. This work bridges the gap between the theoretical understanding of multimodal reasoning and practical implementation considerations. Our future research agenda focuses on two key directions. First, we aim to enhance evaluation methodologies by developing more rigorous test cases and introducing holistic metrics that assess the generalization capabilities of models, ethical alignment, and performance consistency across diverse scenarios. Second, we will advance model performance through innovative approaches, particularly exploring contextual and reinforcement learning techniques to strengthen sophisticated reasoning capabilities across modalities.
	
	% \section*{CRediT authorship contribution statement}
	% \par{Zhangsan: Methodology, Conceptualization, Investigation, Writing - Review \& Editing. }

	% \section*{Declaration of competing interest}
	% \par{The authors declare that they have no known competing financial interests or personal relationships that could have appeared to influence the work reported in this paper.}

	% \section*{Funding}
	% \par{This work was jointly supported by the following projects: }

	% % 致谢
	 \section*{Acknowledgements}
	\par{Thanks to all co-authors for their hard work. The work is supported by the National Natural Science Foundation of China (62172086, 62272092), and the Fundamental Research Funds for the Central Universities under Grant (N25XQD004).}

	\bibliographystyle{model5-names}
	% \bibliographystyle{cas-model2-names}

	\bibliography{MM-InstructEval_references}


	%Author biography without author photo.


\end{sloppypar}
\end{document}